%% file: main.tex
\documentclass[useAMS,usenatbib]{biom}
\usepackage{graphicx}					

\usepackage{amsmath}
\usepackage{amsfonts}				
\usepackage{mathrsfs} 				

\usepackage{bm}					
\usepackage{textcomp}				

\usepackage[hang,small,bf]{caption}
\usepackage[subrefformat=parens]{subcaption}
\usepackage{wrapfig}

\title[Co-Localization Index for Cellular Interactions]{\input{00_title.tex}}

\author{Toru Nagasaka$^{1,2}$\email{toru-ngy@umin.ac.jp} \\
	   1) Association of Medical Artificial Intelligence Curation (AMAIC), Nagoya, Japan
	   \and 
	   Kimihiro Yamashita$^{2}$, Mitsugu Fujita$^{3}$, \\
	   2) Kobe University Hospital, Department of Gastrointestinal Surgery, Kobe, Japan\\
	   3) Kindai University Hospital, Center for Medical Education and Clinical Training, Osaka-sayama, Japan	   
	   }

\begin{document}







\begin{abstract}
\input{00_abstract.tex}
\end{abstract}

%
%


\maketitle

\section{INTRODUCTION}

\input{01_introduction.tex}

\section{THE METHOD}

\input{02_method.tex}

\section{EVALUATION ON SYNTHETIC DATA SETS}

\input{03_simulation.tex}

\section{APPLICATION TO REAL BIOLOGICAL IMAGING}

\input{04_application.tex}

\section{DISCUSSION}

\input{05_discussion.tex}



\section*{Acknowledgements}

We are deeply grateful to Drs. Kuramitsu S, Nakata S, and Ohno M for their extensive proofreading and invaluable contributions to this work. We would also like to express our sincere gratitude to the following individuals for their valuable contributions to the precise image tagging: To the graduate students and technical assistants from Kobe University who assisted with the annotation tasks: Abe T, Adachi Y, Agawa K, Ando M, Fukuda S, Konaka R, Miyake T, Mukoyama T, Okazoe Y, and Ueda Y. We also extend our appreciation to the registered annotators who were engaged by the AMAIC: Adachi K, Akima J, Aoki S, Ichikawa K, Kanto T, Kawase Y, Kimura M, Miura R, Sirasawa H, Sotani K, Suzuki H and Yuki A. Their professional and diligent efforts greatly enhanced the quality of the dataset.
This study is supported by the Grants-in-Aid for Scientific Research from the Ministry of Education, Culture, Sports, Science and Technology of Japan (MEXT; 24K10381 to TN, 23K08171 to KY, and 21K09167 to MF).


\newpage

\bibliographystyle{biom}			%
\bibliography{ref}

\newpage


\input{06_supplement.tex}

\label{lastpage}

\end{document}

%% file: 00_title.tex
Novel Methods for Analyzing Cellular Interactions\break in Deep Learning-Based Image Cytometry: Spatial Interaction Potential and Co-Localization Index

%% file: 00_abstract.tex
The study presents a novel approach for quantifying cellular interactions in digital pathology using deep learning-based image cytometry. Traditional methods struggle with the diversity and heterogeneity of cells within tissues. To address this, we introduce the Spatial Interaction Potential (SIP) and the Co-Localization Index (CLI), leveraging deep learning classification probabilities. SIP assesses the potential for cell-to-cell interactions, similar to an electric field, while CLI incorporates distances between cells, accounting for dynamic cell movements. Our approach enhances traditional methods, providing a more sophisticated analysis of cellular interactions. We validate SIP and CLI through simulations and apply them to colorectal cancer specimens, demonstrating strong correlations with actual biological data. This innovative method offers significant improvements in understanding cellular interactions and has potential applications in various fields of digital pathology.

%% file: 01_introduction.tex
Traditional methods in digital pathology, including optical microscopy-based analysis and standard image processing techniques, struggle to handle cell diversity and heterogeneity. Image analysis plays an essential role and is utilized for measuring various metrics (e.g., nuclear division count, Ki-67 labeling index, HER2 score, etc.). For these measurements, the open-source image processing and analysis programs NIH Image/ImageJ \citep{schneider2012nih} and QuPath are widely used \citep{bankhead2017qupath, humphries2021qupath}. While they are capable of quantifying the tissue structure and characteristics, they often require manual handling due to limitations of automatic processing. Therefore, accurate quantification of the cell diversity and heterogeneity remains a challenge.

Various methods have been introduced into the field of cell-cell interaction analysis.
Recent advances in analyses using optical microscopy have allowed for a better understanding of how cells interact within the complex architecture of tissues \citep{bechtel2021strategies}.
Advances in cell staining have enabled multiplex imaging and brought new techniques to the analysis of cell interactions, such as Image Mass Cytometry \citep{angelo2014multiplexed} and CODEX \citep{goltsev2018deep}.
However, traditional binarization and segmentation methods used in these techniques still struggle to effectively address the dynamics of cell interactions.
Recently, spatial omics analysis has emerged, providing classification and segmentation of cells based on RNA-seq data, to further analyze the cell distribution and interactions.
While this method is used for analyzing cell interactions, it falls under an object-based approach in image analysis.
Despite advancements, there remains a significant challenge in accurately quantifying cellular interactions.

\begin{wrapfigure}{r}[0pt]{0.2\textwidth}
  \includegraphics[width=0.2\textwidth]{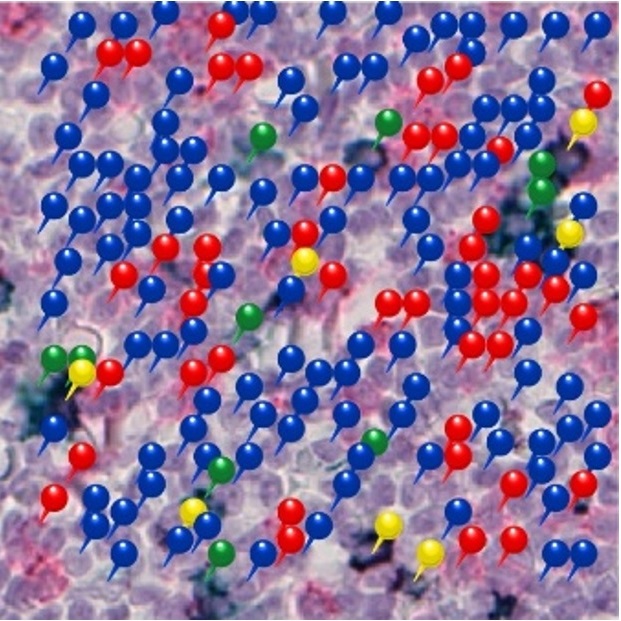}
  \caption{The output of the AI. The probabilities are generated by the softmax function in the final layer of the neural network.}
  \label{fig:ai_out}
\end{wrapfigure}

To analyze cell interactions, it is necessary to address their colocalization. Methods for studying cellular co-localization, which examine the spatial proximity of different cells, can be categorized into two approaches: pixel-based and object-based. Pixel-based methods are highly quantitative, as they directly link signal intensity to cell quantity. Conversely, object-based methods require the initial segmentation of the image into distinct objects \citep{lagache2015statistical}. However, this segmentation process can disrupt the direct relationship between signal intensity and cell quantity.

In recent years, image analysis has been conducted using artificial intelligence (AI) technologies, including deep learning. Since the basic concept of convolutional neural networks (CNNs) was first proposed as LeNet \citep{lecun1998gradient}, numerous improvements and advancements have been made, such as AlexNet \citep{krizhevsky2012imagenet} and ResNet \citep{he2016deep}.

The progress of deep learning methods has led to substantial enhancements in image analysis \citep{jiao2021deep}. A distinct advantage of this approach is its ability to provide “classification probabilities” alongside “labels,” where "labels" refer to the names or identifiers assigned to specific objects or regions in image analysis, and "classification probabilities" represent the likelihood of an object belonging to a particular class or category. Although deep learning-based methods are categorized as object-based methods (refer to the Figure \ref{fig:ai_out}), they enable a more nuanced and probabilistic understanding of object detection.
These advantages open up new possibilities for analysis that were not achievable with traditional object-based methods, which typically involve setting a cutoff value to determine whether an object is present or not, resulting in a binary classification. Nevertheless, detailed reports on methods for co-localization analysis using these new techniques remain limited.

Based on these findings, we hypothesized that deep learning-based classification probabilities represent cell presence and facilitate a nuanced understanding of cellular interactions within complex tissue structures. Before validating this hypothesis, we introduced the novel Spatial Interaction Potential (SIP) and the Cellular Interaction Force (CIF). SIP leverages these classification probabilities to evaluate the potential for cell-to-cell interactions as if in an electric field-like manner, utilizing these distances in CIF to evaluate the strength of interactions that takes into account the dynamics of random cell movements. Furthermore, by defining the Co-Localization Index (CLI) as the sum of CIFs for all cell pair combinations, we enabled the evaluation of spatial interactions between cells. This approach not only exceeds the capabilities of traditional object-based approaches but is also tailored to provide a more sophisticated analysis of the pivotal physical contacts involved in cellular interactions.

The rest of the paper is structured as follows. Section 2 describes the mathematical basis for defining SIP and deriving CLI, and explains the practical considerations made for their implementation. Section 3 demonstrates the results of a simulation study, indicating the efficacy of CLI as a marker of cell interactions. In Section 4, we conduct calculations of SIP and CLI in human clinical samples. Section 5 contains concluding discussions.

%% file: 02_method.tex
\subsection{Mathematical foundations on probability of existence}

This study employed the commercial Cu-Cyto\textsuperscript{\tiny{\textregistered}} automated image analysis platform for its flexibility in integrating supplementary analysis algorithms \citep{abe2023deep}. Although it is proprietary, this platform is based on the universal ResNet architecture \citep{he2016deep}. In the final layer of our model, similar to a conventional network configuration, the platform applies the softmax function to determine cell types (labels) and their associated probabilities. This method allows the model to classify each cell precisely and provide a probability score for each classification.

The probability output from the final layer's softmax function is represented by
\[
\text{softmax}(x)_i = \frac{e^{x_i}}{\sum_j e^{x_j}}
\]

where $x$ is the input vector, and $\text{softmax}(x)_i$ is a specific element of the output vector. The softmax function transforms each element of the input vector into a non-negative value using the exponential function, and subsequently divides each element by the sum of these values to represent them as probabilities. Due to the properties of the softmax function, each element of the output falls within the range of 0 to 1, with their total sum equaling 1. Therefore, $\text{softmax}(x)_i$ can be interpreted as indicating the probability that $x$ belongs to the $i$th class (or category).

Treating the classification probabilities as probabilities of cell presence, the output probabilities can be represented as conditional probabilities learned under specific training sets. The learning algorithm aims to train a model to discover a function $f: x \rightarrow y$ that predicts target $y$ from input data $x$. With a collection of training datasets $D = \{(x_1, y_1), (x_2, y_2), \ldots, (x_N, y_N)\}$, a model $M$ trained on $D$ outputs a probability $P_M(y=i|x,D)$ for an input $x$, indicating the probability that $x$ belongs to class $i$. This probability is estimated by the model $M$ based on the training dataset $D$. Therefore, the conditional probability $P_M(y=i|x,D)$ that $x$ belongs to class $i$ should closely approximate the existence probability $p_i$ if the size $ |D| $ of the training dataset is sufficiently large.
\[
\lim_{{|D| \to \infty}} P_M(y=i|x,D) \approx p_i
\]
This approximation is valid when the training dataset $D$ effectively reflects the true population $P$. That is, it hinges on whether the training dataset has been created with precise and comprehensive information by annotating all identifiable cell types during cell annotation. Let us denote the accuracy of this annotation as probability $q$.
\[
\lim_{{|D| \to \infty}} P_M(y=i|x,D) = p_i \quad \land \quad D \overset{q}{\rightarrow} P
\]
If the training dataset is sufficiently large and accurately reflects the population, it is reasonable to treat classification probabilities as existence probabilities. In other words, if the training dataset $D$ asymptotically reflects the true population $P$ under the probability $q$, it is reasonable to treat classification probabilities as existence probabilities.

\subsection{Introduction of spatial interaction potential}

In pathological specimens, cells are immobilized due to processes such as formalin fixation, although it is presumed that they were capable of movement prior to fixation. This limitation in temporal information can hinder the accurate assessment of contact events. To address this, consider a cell at the origin $(0,0)$ with a probability of presence denoted as $p$, and an interaction constant $K_c$. The Spatial Interaction Potential (SIP) $\Phi$ at a coordinate $(x, y)$ can be defined as follows:
\[
\Phi = K_c \frac{p}{x^2 + y^2}
\]
This formulation allows for the definition of a field of cellular interactions, analogous to magnetic or electric fields, facilitating the consideration of potential contact scenarios.
The spatial interaction potential decays proportionally to the square of the distance $r^2$ from the cell's position. This potential is thought to represent the distribution of the likelihood of neighboring cells having either previously come into contact with the cell or potentially making contact with it in the future, under the assumption that the cell remains stationary.

\begin{wrapfigure}{r}[0pt]{0.2\textwidth}
  \centering
  \includegraphics[width=0.2\textwidth]{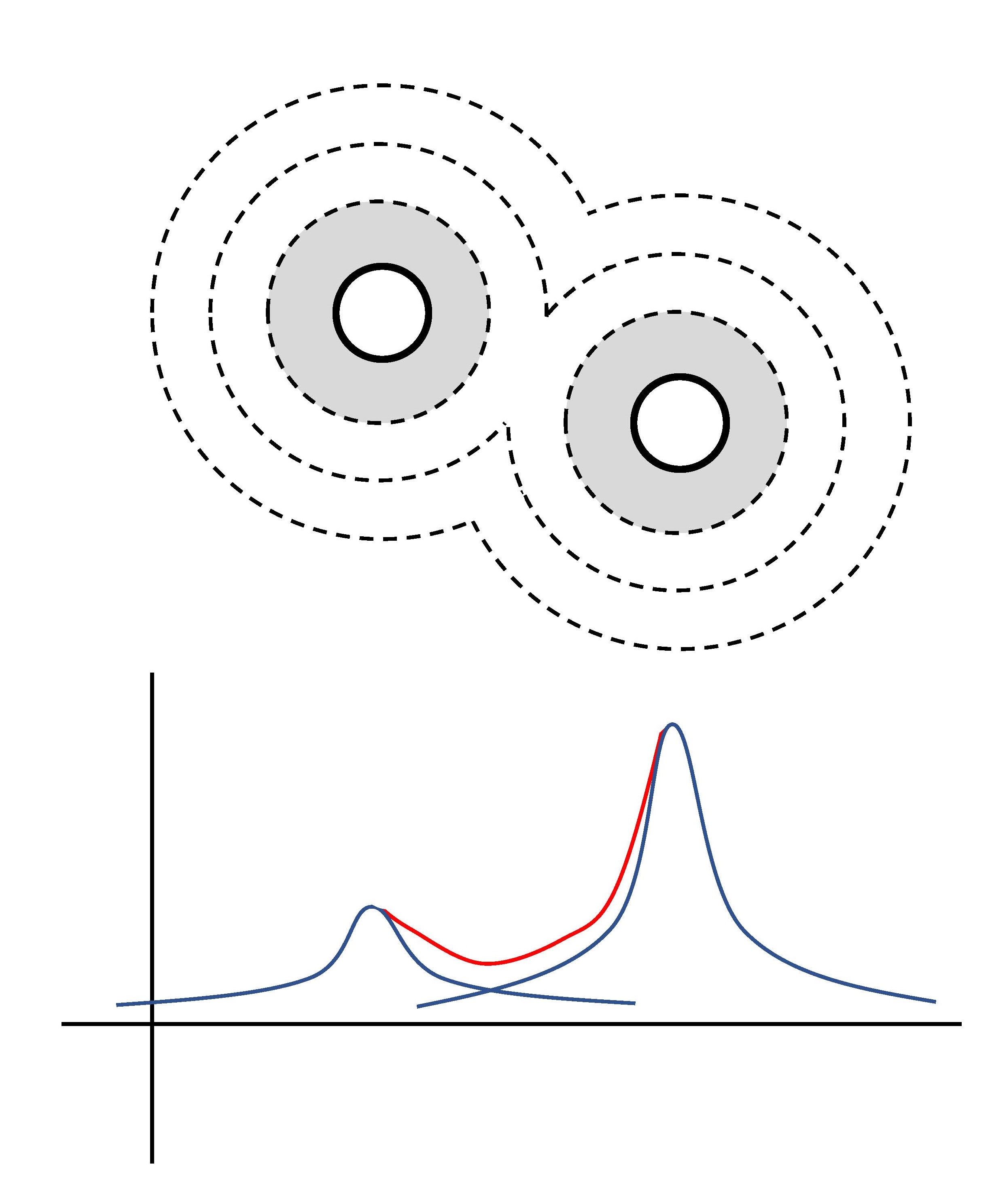}
  \caption{Space interaction potential}
  \label{fig:phi_fig}
\end{wrapfigure}

Let us consider the case where two cells are present. Around the cell denoted as A, the previously defined cellular interaction potential is distributed around it. Similarly, around the cell denoted as B, a similar potential is distributed. These cells, A and B, have the potential to have come into contact in the past or may potentially come into contact in the future. In such a scenario, it is conceivable that the possibility of cells A and B coming into contact at any given coordinate can be compounded.

The composite potential of the two types of spatial interaction potentials, denoted as $\Phi_c$, is defined as follows. The first potential is denoted as $\Phi_1$, and its value at the position coordinates $(x, y)$ is denoted as $\Phi_1(x, y)$. Similarly, the second potential is denoted as $\Phi_2$, and its value at the position coordinates $(x, y)$ is denoted as $\Phi_2(x, y)$. The composite potential $\Phi_c$ is the sum of $\Phi_1(x, y)$ and $\Phi_2(x, y)$ (Figure \ref{fig:phi_fig}). The composite spatial interaction potential $\Phi_c$ can be visualized at each position coordinate $(x, y)$, and this is referred to as a spatial interaction map. Using this spatial interaction map, we conducted the visualization of cell-cell interactions and the evaluation of CLI.
\[
\Phi_c(x, y) = \Phi_1(x, y) + \Phi_2(x, y)
\]

In practice, a grid with intervals of 3.843$\mu$m is set up, and the synthetic spatial interaction potential $\Phi_c$ is calculated at each intersection coordinate. It is possible to calculate $\Phi_c$ for any combination of two or more types of cells.

\subsection{Spatial cellular interaction}

\begin{wrapfigure}{r}[0pt]{0.2\textwidth}
  \centering
  \includegraphics[width=0.2\textwidth]{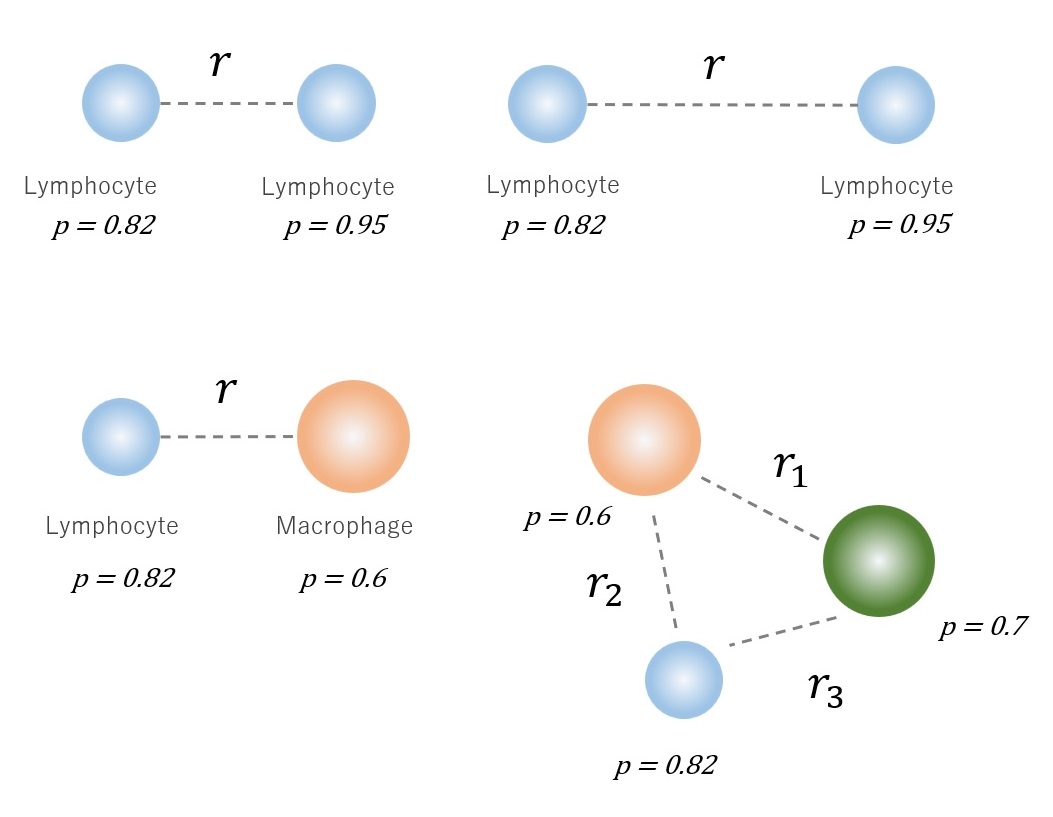}
  \caption{Cellular space interaction}
  \label{fig:cell_inter}
\end{wrapfigure}

The Cellular Interaction Force (CIF), denoted as $\Psi$, can be derived from the spatial interaction potential $\Phi$ by considering the presence probabilities of cells 1 and 2 as $p_1$ and $p_2$, respectively, along with the cell-cell distance $r$ and the interaction constant $K_c$ (see Figure \ref{fig:cell_inter}). The relationship is given by the equation:
\[
\Psi = K_c \frac{p_1 p_2}{r^2}
\]
This expression demonstrates that $\Psi$ is directly proportional to the product of the presence probabilities $p_1$ and $p_2$, and inversely proportional to the square of the cell-cell distance $r$. Notably, this relationship holds true for both cells of the same type and of different types.

\subsection{Defining co-localization index for cellular interaction}

\begin{figure}[h]
  \centering
  \includegraphics[width=6.0cm]{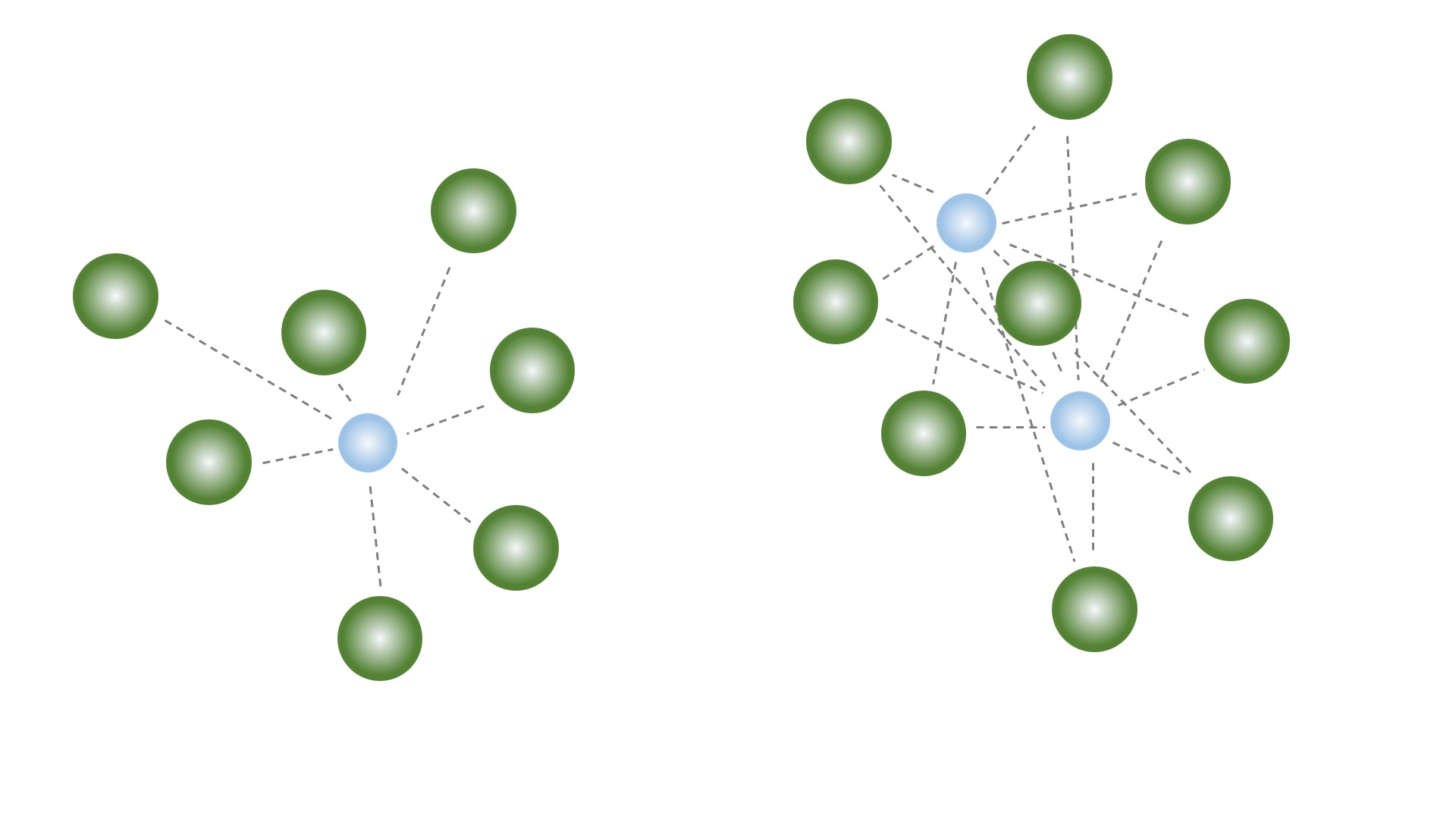}
  \caption{Co-localization index}
  \label{fig:CLI}
\end{figure}

Leukocytes are known to cluster via chemotaxis, activate through cell-cell contacts, and proliferate through various stimuli. In actual cellular microenvironments, complex interactions involving multiple-to-multiple interactions are observed, indicating that the cellular microenvironment is composed of intricate interaction networks. To account for such circumstances and introduce indicators based on cell density and distance, it is imperative to consider the cumulative summation of cell-cell spatial interactions, denoted as the total cell-cell spatial interaction $\Psi$.
In order to quantitatively assess densely interacting cells, we have defined the Co-Localization Index (CLI) between two cell populations, A and B, as the sum of all cell-cell spatial interactions $\Psi$ \citep{nagasaka2021cli_patent}. The CLI between cells in the two populations, A and B, is defined as follows (Figure \ref{fig:CLI}):
\[
\Omega = K_c \sum_{i\in A} \sum_{j\in B} \frac{\bm{p}_i \bm{p}_j}{\bm{r}^2}
\]

If a single type of cell belongs to a cell set $A$ consisting of $n$ elements, the co-localization index $\Omega_1$ is given by:
\[
\Omega_1 = K_c \sum_{1 \leq i < j \leq n} \frac{\bm{p}_i \bm{p}_j}{\bm{r}^2}
\]

\subsection{Extending co-localization index to multiple cell types}

\begin{figure}[h]
  \centering
  \includegraphics[width=8.0cm]{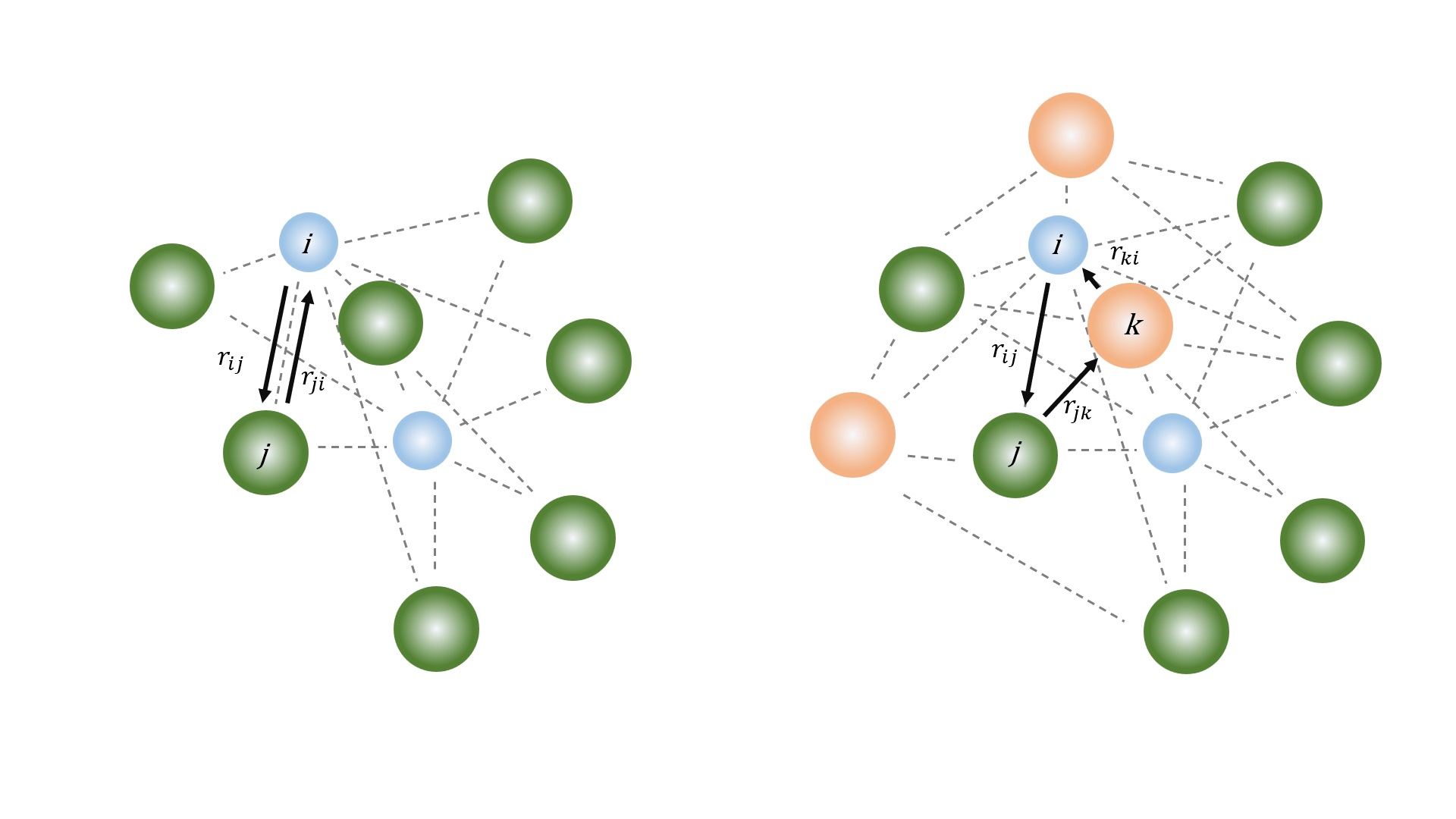}
  \caption{The co-localization index for 2 or 3 types of cell interactions}
  \label{fig:CLIN}
\end{figure}

The square of the distance between two cell types, denoted as $\bm{r}^2$, is generalized as the product of the distances $r_{ij}$ between cells $i$ and $j$ and $r_{ji}$ between cells $j$ and $i$. Similarly, this suggests the extensibility of the co-localization index for $N$ types of cell interactions, as illustrated in Figure \ref{fig:CLIN}.

\begin{align}
\Omega_2 &= K_c \sum_{i \in A} \sum_{j\in B} \frac{\bm{p}_i \bm{p}_j}{\bm{r}^2} \nonumber \\
&= K_c \sum_{i\in A} \sum_{j\in B} \frac{\bm{p}_i \bm{p}_j}{r_{ij} r_{ji}} \nonumber \\
\nonumber \\
\Omega_3 &= K_c \sum_{i\in A} \sum_{j\in B} \sum_{k\in C} \frac{\bm{p}_i \bm{p}_j \bm{p}_k}{r_{ij} r_{jk} r_{ki}} \nonumber \\
\nonumber \\
\Omega_4 &= K_c \sum_{i\in A} \sum_{j\in B} \sum_{k\in C} \sum_{l\in D} \frac{\bm{p}_i \bm{p}_j \bm{p}_k \bm{p}_l}{r_{ij} r_{jk} r_{kl} r_{li}} \nonumber
\end{align}

\begin{figure}[h]
  \centering
  \includegraphics[width=8.0cm]{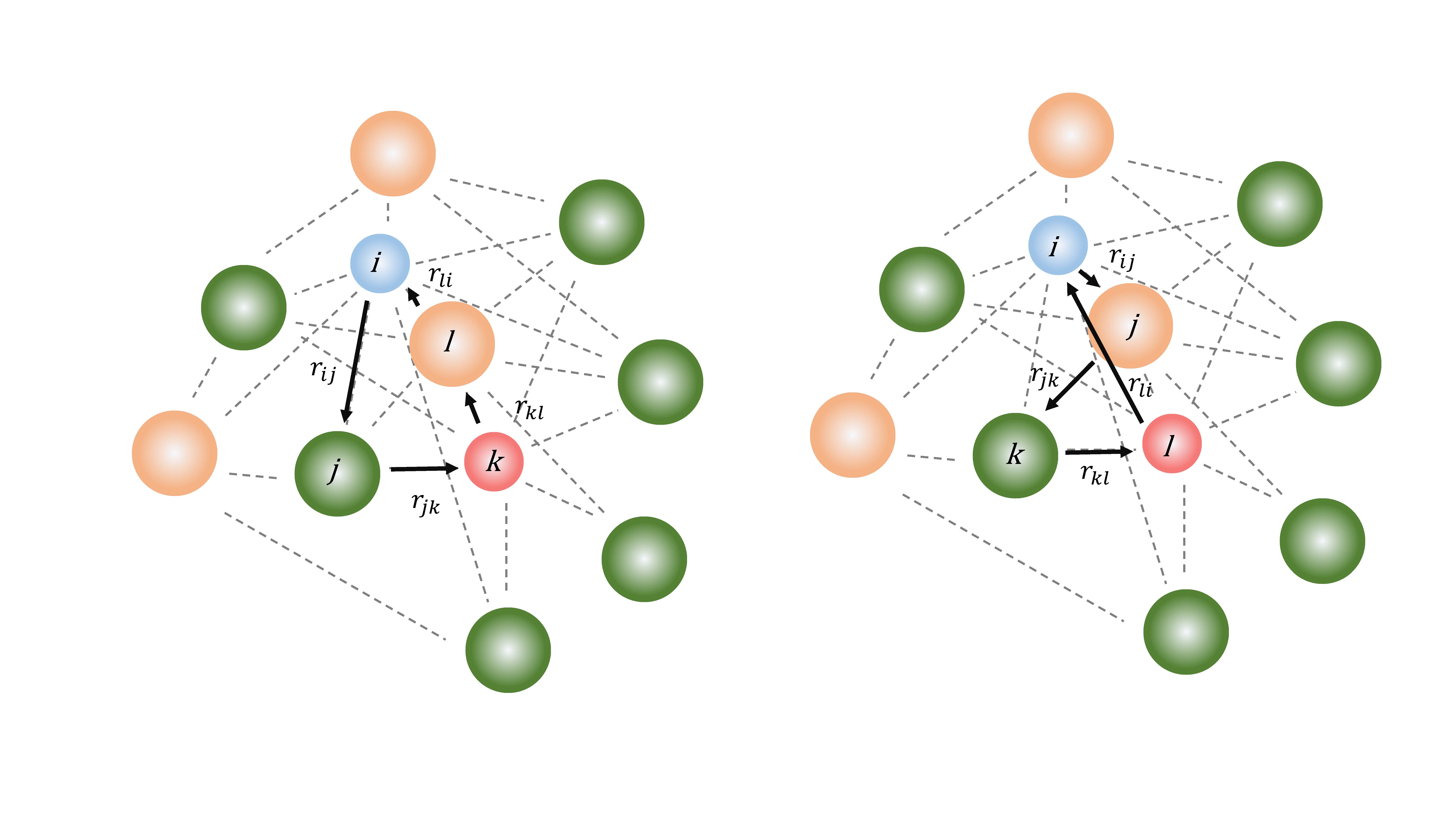}
  \caption{The co-localization index for N types of cell interactions}
  \label{fig:CLIgreq4}
\end{figure}

However, for $N \ge 4$, the calculation of distances is not uniquely determined due to the ambiguity in the path selection. Consequently, the co-localization index is not uniquely defined for such cases. Nonetheless, it is feasible to compute the CLI by specifying the specific order of cells (Figure \ref{fig:CLIgreq4}).

\subsection{Optimizing tile patterns for cell packing and standardization of cell detection intervals}

\begin{figure}[h]
 \centering
 \begin{minipage}[b]{0.45\linewidth}
  \centering
  \includegraphics[width=0.5\linewidth]{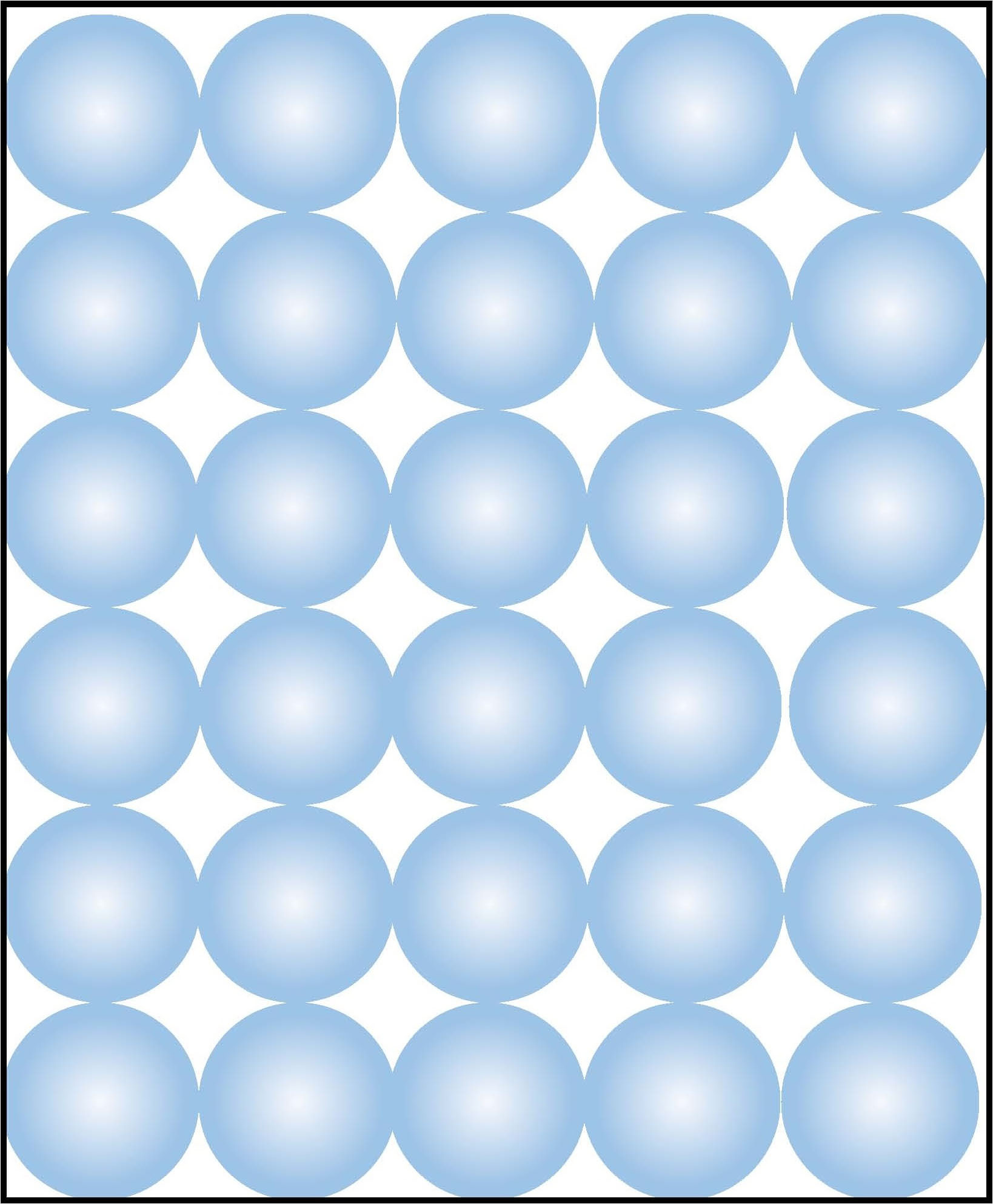}
  \subcaption{Lattice cell array}
  \label{fig:tilingA}
 \end{minipage}
 \begin{minipage}[b]{0.45\linewidth}
  \centering
  \includegraphics[width=0.5\linewidth]{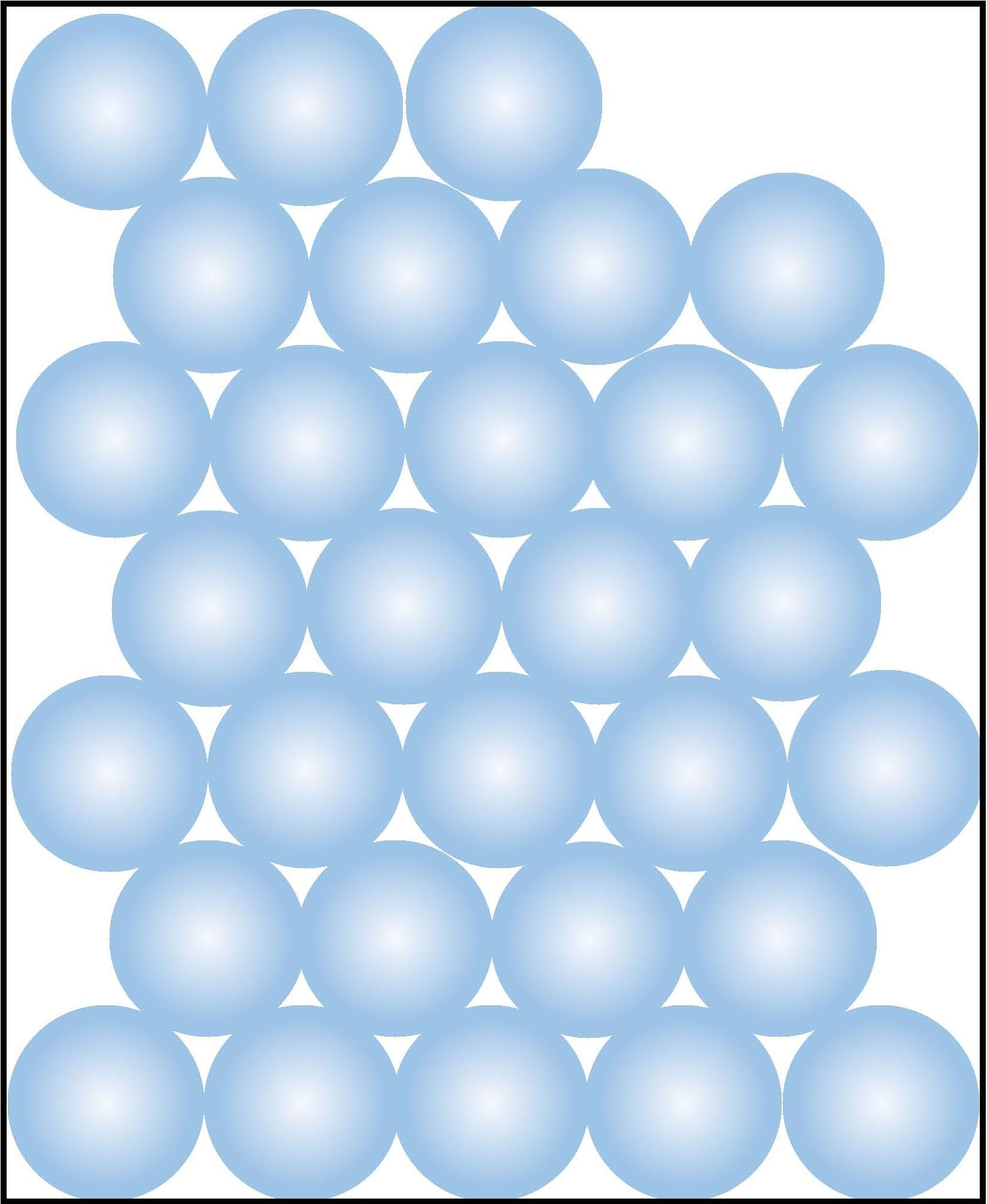}
  \subcaption{Compact cell array}
  \label{fig:tilingB}
 \end{minipage}
  \caption{Cell array}
\end{figure}

Tiling refers to the process of covering a specified region on a plane with squares or regular polygons without leaving any gaps. Mathematically, the tiling problem involves determining whether it is possible to cover a given area without overlapping tiles \citep{conway1990tiling}. The tiling of circles, known as the "coin packing problem," is a well-known challenge. It has been established that arranging the circles alternately (Figure \ref{fig:tilingB}) allows for more efficient packing of coins compared to a regular grid pattern (Figure \ref{fig:tilingA}). While considering densely packed cell arrangements, the ideal scenario to consider would be a configuration akin to Figure \ref{fig:tilingB}. However, for computational simplicity, we chose to examine the densely packed lattice-like arrangement depicted in Figure \ref{fig:tilingA}.

The cell diameter of 3.843 $\mu$m was intentionally selected as the scanning interval for our cell detection, to assume sizes smaller than lymphocytes, which are among the smallest types of cells. While the selection of this constant may involve some degree of subjectivity, the approach of setting the baseline to 1 based on the densely and regularly arranged small cells is considered to contribute to the standardization of measurement results across different cases. Similarly, it is inferred that a certain standardization has been employed in the comparison between single-cell CLI, dual-cell CLI, and triple-cell CLI.

\subsection{Determination of cellular interaction constant $Kc$}

\begin{figure}[h]
 \centering
 \begin{minipage}[b]{0.45\linewidth}
  \centering
  \includegraphics[width=0.6\linewidth]{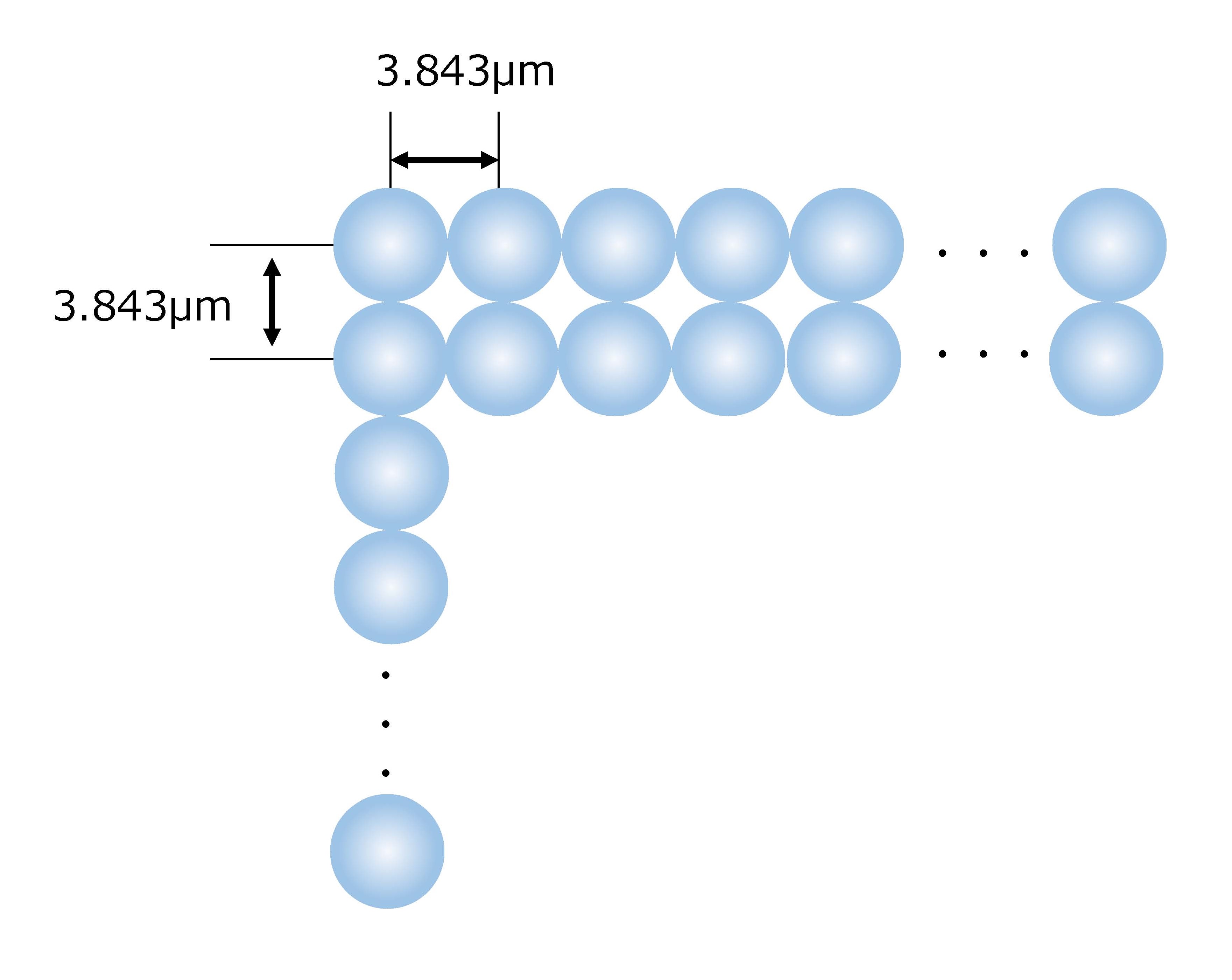}
  \subcaption{Compact arrangement of a single cell type}
  \label{fig:Kc1}
 \end{minipage}
 \begin{minipage}[b]{0.45\linewidth}
  \centering
  \includegraphics[width=0.8\linewidth]{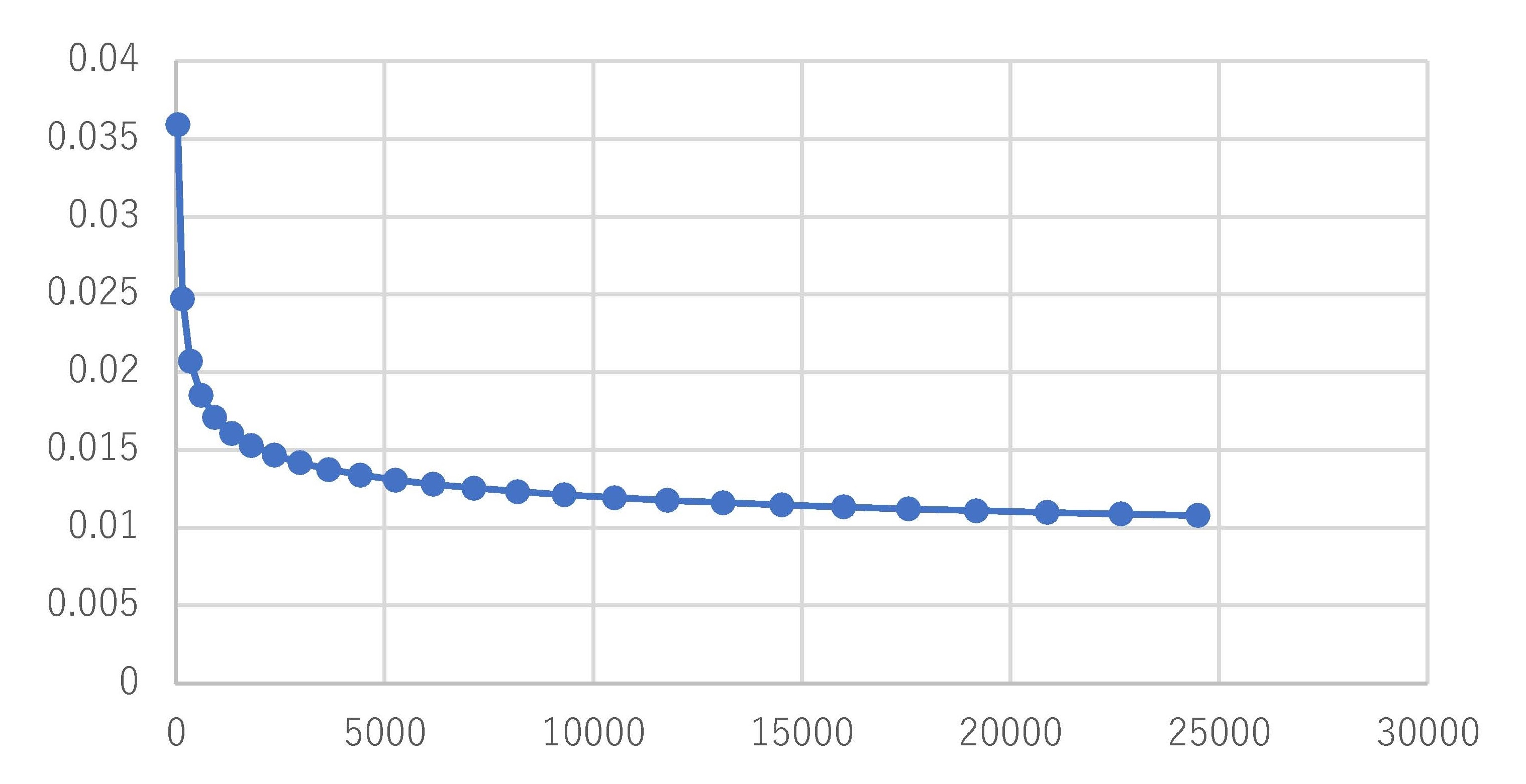}
  \subcaption{$Kc_1$}
  \label{fig:Kc1graph}
 \end{minipage}
 \caption{Cellular interaction constant $Kc_1$ of a single cell type}
\end{figure}

The constant $K_{c}$ is defined as the correction factor to ensure that the CLI per 100 cells is 1.0 when cells with a diameter of 3.843 $\mu$m and a presence probability of 1.0 are arranged in a lattice pattern (see Figure \ref{fig:Kc1}). According to this definition, as $k \to \infty$, $K_{c1}$ does not converge but approaches $0$. However, for practical reasons, the computation was terminated at 8 significant figures, where the calculation results had stabilized. Following this approach, $K_{c1}$ was determined to be $0.010781739$ (see Figure \ref{fig:Kc1graph}).

\begin{figure}[h]
 \centering
 \begin{minipage}[b]{0.45\linewidth}
  \centering
  \includegraphics[width=0.6\linewidth]{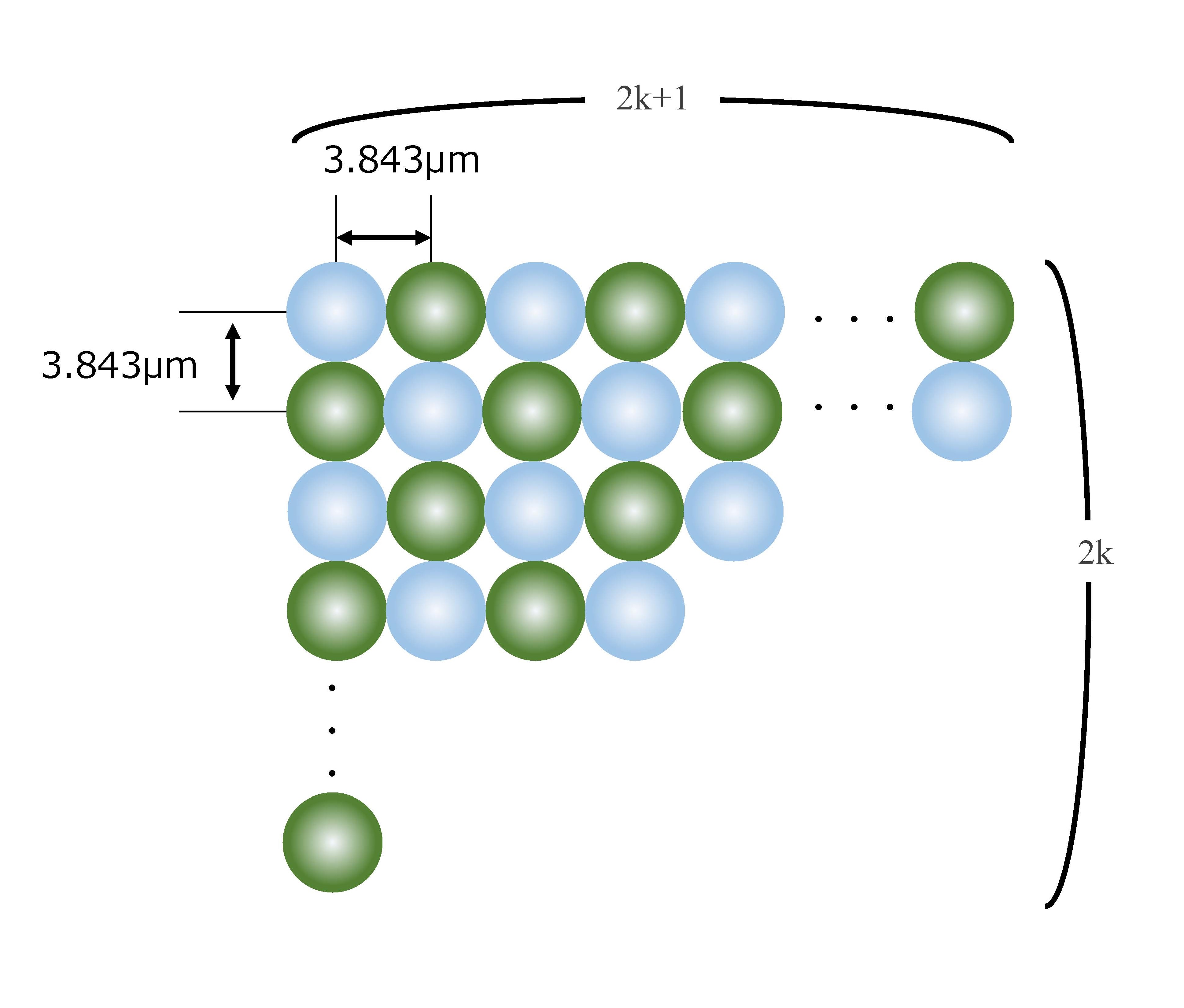}
  \subcaption{Co-localization indices for two types of cells}
  \label{fig:Kc2}
 \end{minipage}
 \begin{minipage}[b]{0.45\linewidth}
  \centering
  \includegraphics[width=0.8\linewidth]{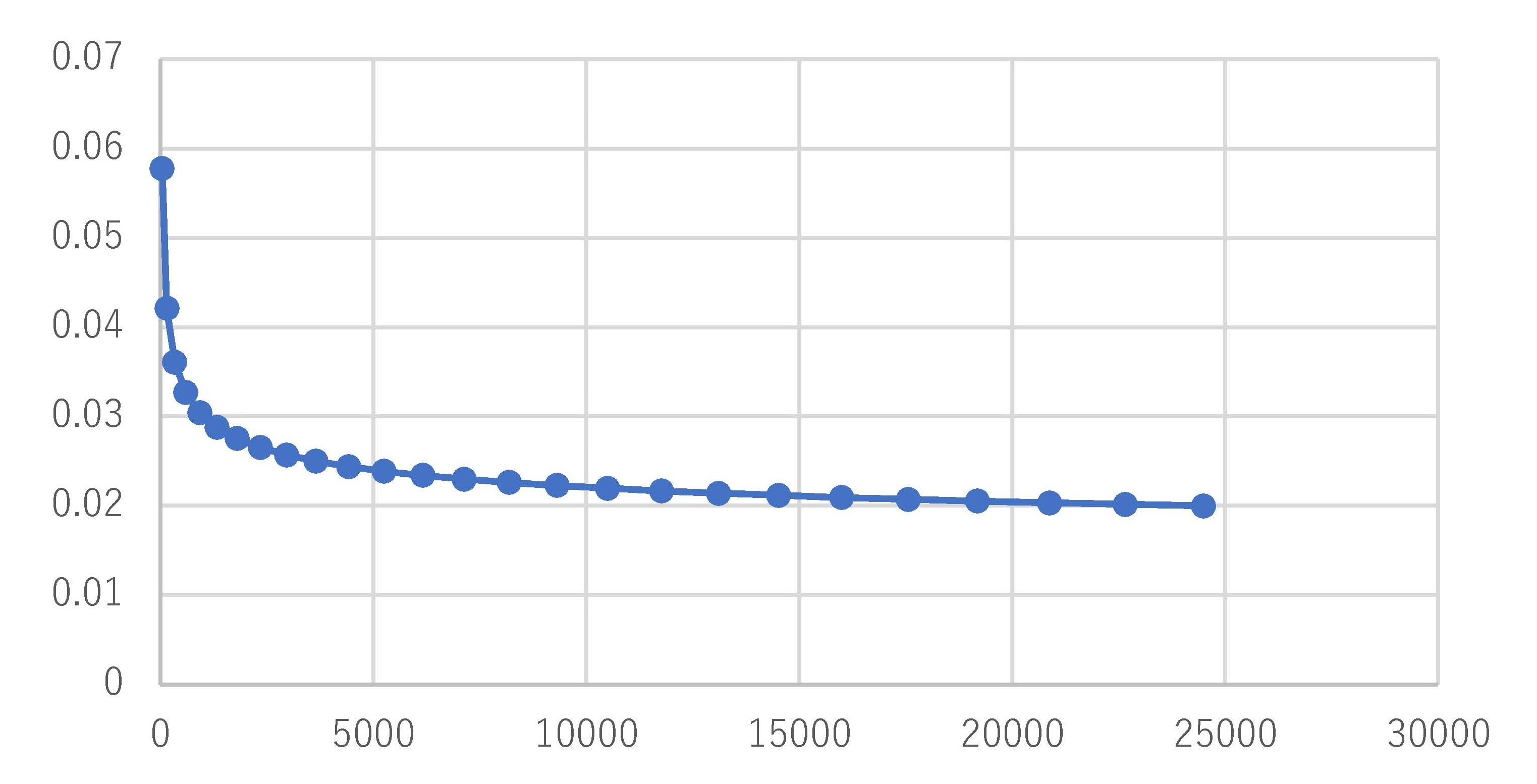}
  \subcaption{$Kc_2$}
  \label{fig:Kc2graph}
 \end{minipage}
 \caption{Cellular interaction constant for two types of cells $Kc_2$}
\end{figure}

Next, we examined the interaction constant $Kc_2$ between two types of cells. We assumed an ideal state where the two types of cells are alternately arranged in a regular pattern (see Figure \ref{fig:Kc2}). We created a list of $(2k+1) \times 2k$ cells as follows:
${\scriptsize
\begin{bmatrix}
1 & 2 & 1 & 2 & ... & 2\\
\end{bmatrix}
}$
and arranged it into a matrix of size $2k \times (2k+1)$ as follows:
${\tiny
\begin{bmatrix}
1 & 2 & 1 & 2 & 1 & \cdots & 1 \\
2 & 1 & 2 & 1 & 2 &  &  2 \\
1 & 2 & 1 & 2 & 1 &  &  1  \\
2 & 1 & 2 & 1 & 2 &  &  2 \\
1 & 2 & 1 & 2 & 1 &   & 1 \\
\vdots &    &    &    &  & \ddots &  \vdots \\
2 &1 & 2 & 1 & 2 &  \cdots & 2 
\end{bmatrix}}$

The CLI between the two cells was calculated in a similar manner. As we increased $k$, the CLI per 100 cells approached a constant value, resulting in $Kc_2=0.019980203$ (see Figure \ref{fig:Kc2graph}).

For three cells, similarly, we considered an ideal scenario where three types of cells are regularly arranged in an alternating fashion. We created a list of $(3k+1) \times 3k$ cells as follows:
$
{\scriptsize
\begin{bmatrix}
1 & 2 & 3 & 1 & 2 & 3 & 1 & \ldots & 3\\
\end{bmatrix}
}
$
and rearranged it into a $3k \times (3k+1)$ matrix as shown below:
$
{\tiny
\begin{bmatrix}
1 & 2 & 3 & 1 & 2 & \cdots & 1 \\
2 & 3 & 1 & 2 & 3 & & 2 \\
3 & 1 & 2 & 3 & 1 & & 3 \\
1 & 2 & 3 & 1 & 2 & & 1 \\
2 & 3 & 1 & 2 & 3 & & 2 \\
\vdots & & & & & \ddots & \vdots \\
3 &1 & 2 & 3 & 1 & \cdots & 3
\end{bmatrix}
}
$
Consequently, we obtained $Kc_3=0.007303239$.

\begin{figure}[h]
 \centering
 \begin{minipage}[b]{0.45\linewidth}
  \centering
  \includegraphics[width=0.6\linewidth]{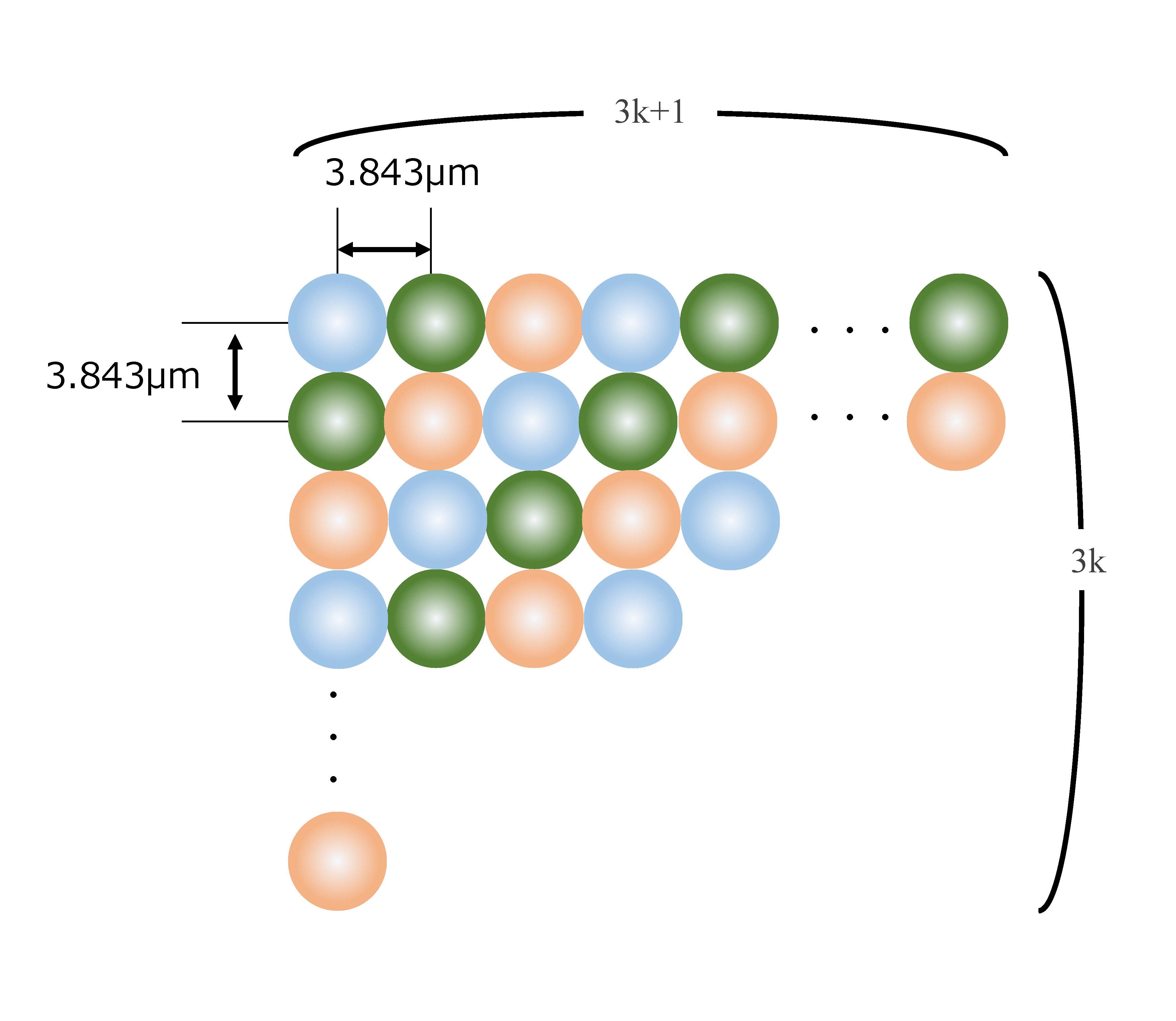}
  \subcaption{Compact arrangement of three types of cell}
  \label{fig:Kc3}
 \end{minipage}
 \begin{minipage}[b]{0.45\linewidth}
  \centering
  \includegraphics[width=0.8\linewidth]{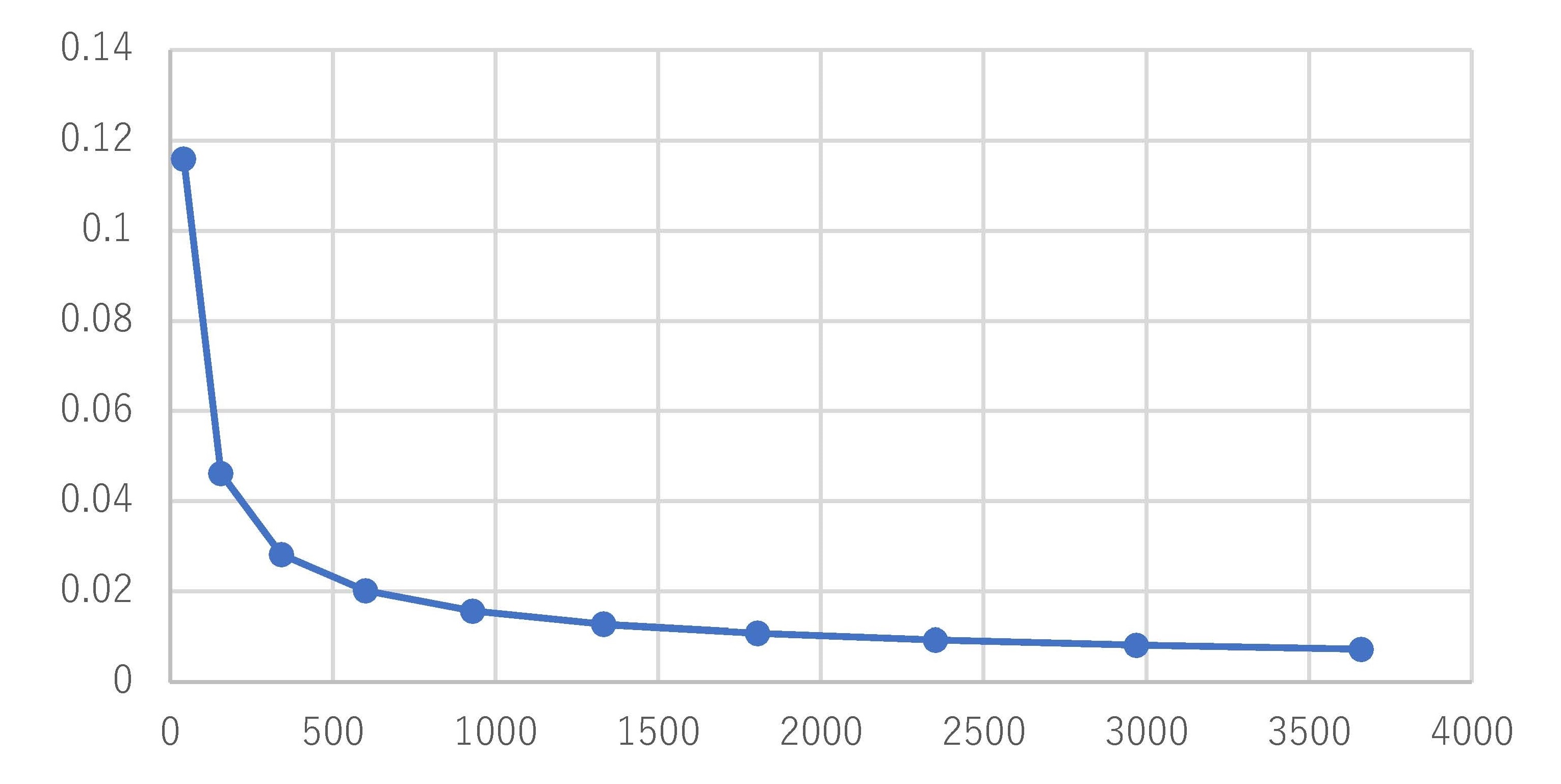}
  \subcaption{$Kc_3$}
  \label{fig:Kc3graph}
 \end{minipage}
 \caption{Cellular interaction constant for three types of cells $Kc_3$}
\end{figure}

\subsection{Materials and Application for actual biological images}

\subsubsection*{Patients}
\hspace{20pt}

The study protocol received approval from the Institutional Review Board at Kobe University in Kobe, Japan (approval number: B200244); the requirement for written informed consent was waived due to the retrospective design of this study. Information about this study was made publicly available on our center's website. Patient confidentiality was rigorously protected according to ethical guidelines. We performed retrospective histopathological analyses of archived formalin-fixed, paraffin-embedded surgical samples from 70 patients diagnosed with colorectal cancer, who had undergone radical surgery at our institution between January 2005 and December 2016. Institutional pathologists conducted the histopathological examination and diagnosis.

\subsubsection*{Tissue Immunohistochemical Staining}
\hspace{20pt}

The procedures used in this study have been previously described \citep{abe2023deep}. Briefly, tissue sections of 4-$\mu m$ thickness underwent the following staining process. These sections were initially deparaffinized with xylene and rehydrated through a series of ethanol dilutions. Antigen retrieval was performed by heat-induced epitope retrieval. Subsequently, sections were subject to immunostaining using the following primary monoclonal antibodies (mAbs): an anti-CD8 mouse mAb (clone C8/144B, DAKO, Glostrup, Denmark), an anti-CD103 rabbit mAb (clone EPR4166(2), abcam). Dual staining for CD8 and CD103 utilized Histofine Simple Stain AP (M) (mouse, Nichirei) and Histofine Simple Stain MAX-PO (R) (rabbit, Nichirei) as secondary antibody complexes. Chromogenic development was achieved using First Red II substrate kit (Nichirei) and HistoGreen substrate kit (Eurobio-Abcys, Les Ulis, France). The stained tissue sections were digitally scanned using a NanoZoomer-SQ whole slide imaging system (Hamamatsu Photonics, Hamamatsu, Japan) with a 20 $ \times $ 0.75 NA objective lens, according to standardized protocols.
From the patch images containing both cancer and lymphocytes, 286 images were randomly selected. Cells were detected using Cu-Cyto\textsuperscript{\tiny{\textregistered}}, and both CLI and SIP were calculated.

This chapter introduces a spatial interaction map for visualizing SIP between cells and defines the CLI for evaluating cell-cell interactions. Leveraging the individual probabilities of cell presence, the CLI serves as a crucial metric for assessing potential interactions based on cell-to-cell distances. Additionally, Section 3 presents simulations that show how the cumulative number of cell contacts changes with the initial distance $r$ between two cells, demonstrating that an inverse proportionality to $r^2$ is a valid assumption for these interactions. Furthermore, in Section 4, we use actual biological images to calculate CLI within the target region and examine the correlation between CLI and the sum of synthetic SIP $\sum \Phi_c$ at every point of a grid-based coordinate system.

%% file: 03_simulation.tex
\subsection{Modeling cell movement and contact frequency using one-demensional random walk}

\begin{wrapfigure}{r}[0pt]{0.2\textwidth}
  \centering
  \includegraphics[width=0.2\textwidth]{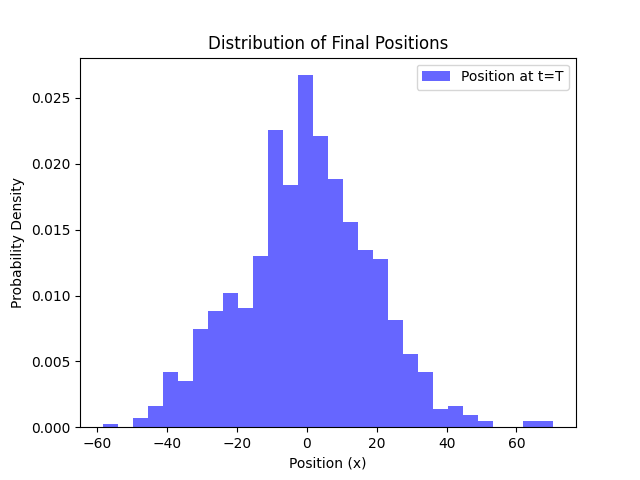}
  \caption{Simulation of cell positions in a random walk model}
  \label{fig:cell_position}
\end{wrapfigure}

In this section, we aim to investigate the effectiveness of the previously defined CLI as an indicator of cell-cell interactions. To achieve this, we conducted simulations to observe variations in the actual number of cell contacts based on cell-to-cell distances. First, we adopted a one-dimensional random walk model for simplicity. In this model, the cell's position $X(t)$ at time $t$ is described as the cumulative positional change from its initial position, given by:
\[
X(t) = O + \sum_{i=1}^{t} \Delta x_i
\]

Here, $\Delta x_i$ represents the change in position at time step $i$ and is a random variable with a probability distribution depending on the specific situation of the random walk, including uniform, normal, exponential, and power-law distributions.
Although lymphocyte migration is influenced by factors such as leukocyte chemotaxis, we adopt a uniform distribution here for simplicity. The simulation results of cell positions after a certain period at specific time $T$ demonstrates a normal distribution centered around the original origin (Figure \ref{fig:cell_position}).
The simulation program can be found in Supplement 1.

Next, we conduct a simulation counting the number of random contacts between two cells in a one-dimensional space. As mentioned earlier, the position $X_1(t)$ of cell 1 at the origin initially is described as follows:
\[
X_1(t) = O + \sum_{i=1}^{t} \Delta x_i
\]
Then, we define the position $X_2(t)$ of cell 2 at an initial coordinate $r$ as
\[
X_2(t) = r + \sum_{i=1}^{t} \Delta x_i
\]
We consider the distance between the cells, $d(t) = \left| X_2(t) - X_1(t) \right|$, and define the cumulative contact function as follows when the distance is less than a fixed distance $d_0$:
\[
C(r,t) = \sum_{t=1}^{T} \delta(d(i) < d_0)
\]

Here, $d(t)$ represents the distance at time $t$, $t_i$ represents the time step $i$, and $\delta(\cdot)$ is the Dirac delta function that returns 1 when the condition is true and 0 otherwise. If the condition $r(t_i) < d$ is true, a contact occurs and 1 is counted.
The simulation program can be found in Supplement 2.

\begin{figure}[h]
  \centering
  \includegraphics[width=0.3\textwidth]{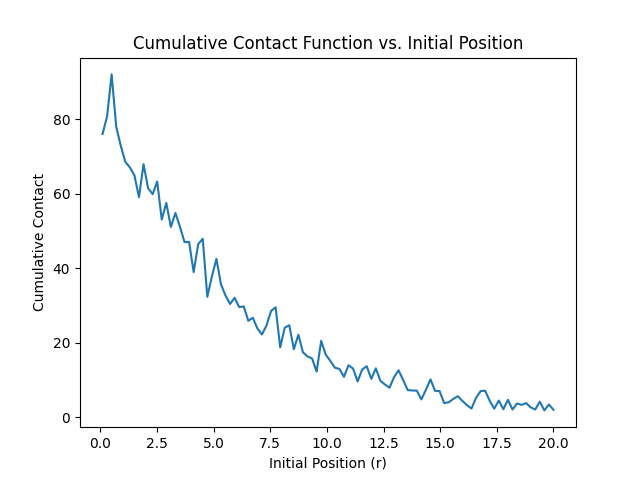}
  \caption{Simulation of cumulative contact counts in one-dimensional random walk model}
  \label{fig:cumurative_01}
\end{figure}


\subsection{Two-demensional random walk model}

\begin{wrapfigure}{r}[0pt]{0.2\textwidth}
  \centering
  \includegraphics[width=0.2\textwidth]{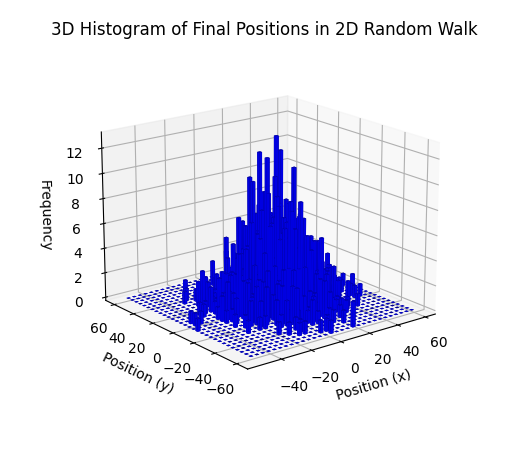}
  \caption{Simulation of cell positions in two-dimensional random walk model}
  \label{fig:cell_position2}
\end{wrapfigure}

Next, we conducted simulations using a two-dimensional random walk model. Similar to the one-dimensional case, we present the simulation results of cell positions at time $T$ after a certain period (Figure \ref{fig:cell_position2}). The cell positions after a fixed period conform to a normal distribution centered around the origin. The simulation program can be found in Supplement 3.

Next, we conducted simulations to count the number of encounters between two cells randomly moving in two dimensions. Random walks along the $x$ and $y$ axes were generated from a uniform distribution, and the cumulative contact count was determined based on the cell-to-cell distances. Here, $\Delta x_i$ and $\Delta y_i$ represent the changes in the $x$ and $y$ positions at time step $i$, respectively, and are random variables. The $x$ and $y$ coordinates of cell 1 at time $t$, $X_1(t)$ and $Y_1(t)$, were as follows:
\begin{align}
X_1(t) = O + \sum_{i=1}^{t} \Delta x_i \nonumber \\
Y_1(t) = O + \sum_{i=1}^{t} \Delta y_i \nonumber
\end{align}

Similarly, the positions $X_2(t)$ and $Y_2(t)$ of cell 2, initially located at coordinate $r$, were defined as:

\begin{align}
X_2(t) = r + \sum_{i=1}^{t} \Delta x_i \nonumber \\
Y_2(t) = O + \sum_{i=1}^{t} \Delta y_i \nonumber
\end{align}

\begin{figure}[h]
  \centering
  \includegraphics[width=0.3\textwidth]{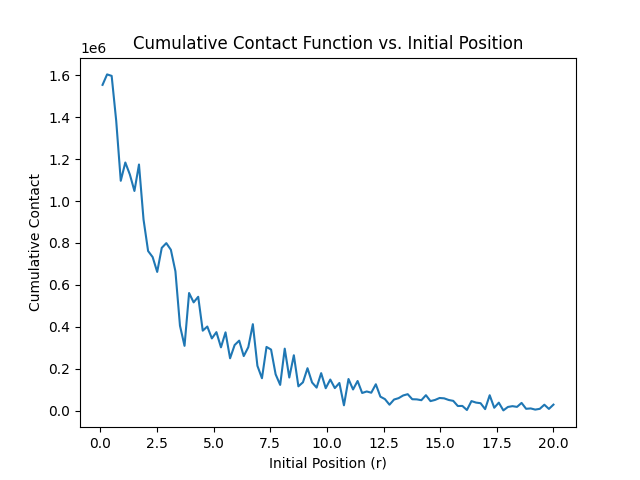}
  \caption{Simulation of cumulative contact counts in two-dimensional random walk model}
  \label{fig:cumurative_02}
\end{figure}

We plotted the contact count $C(r,t)$ with varying initial cell-to-cell distances $r$ (Figure \ref{fig:cumurative_02}). As the initial cell-to-cell distance $r$ increased, the cumulative contact count decreased. The simulation program can be found in Supplement 4.


\subsection{Three-demensional random walk model}

\begin{wrapfigure}{r}[0pt]{0.2\textwidth}
  \centering
  \includegraphics[width=0.2\textwidth]{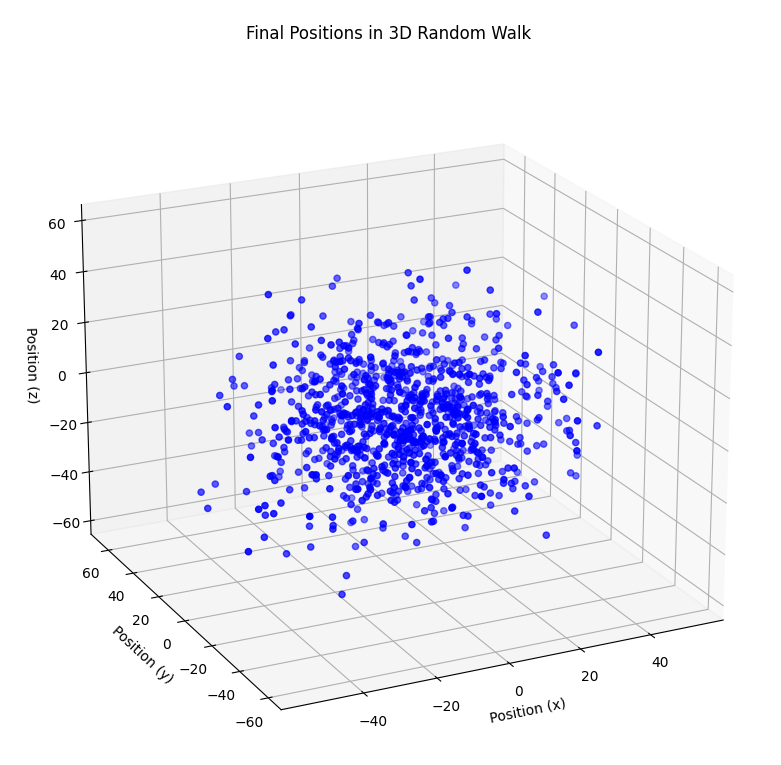}
  \caption{Simulation of cell positions in three-dimensional random walk model}
  \label{fig:cell_position3}
\end{wrapfigure}

Next, we conducted simulations using a three-dimensional random walk model. Similar to the one-dimensional and two-dimensional case, we present the simulation results of cell positions at time $T$ after a certain duration (Figure \ref{fig:cell_position3}). The cell positions after a fixed period conform to a normal distribution centered around the origin. The simulation program can be found in Supplement 6.

Next, we conducted simulations to count the number of encounters between two cells randomly moving in three dimensions. Random walks along the $x$, $y$ and $z$ axes were generated from a uniform distribution, and the cumulative contact count was determined based on the cell-to-cell distances. Here, $\Delta x_i$, $\Delta y_i$ and $\Delta z_i$ represent the changes in the $x$,  $y$ and $z$ positions at time step $i$, respectively, and are random variables. The $x$, $y$ and $z$ coordinates of cell 1 at time $t$, $X_1(t)$, $Y_1(t)$ and $Z_1(t)$, were as follows:
\begin{align}
X_1(t) = O + \sum_{i=1}^{t} \Delta x_i \nonumber \\
Y_1(t) = O + \sum_{i=1}^{t} \Delta y_i \nonumber \\
Z_1(t) = O + \sum_{i=1}^{t} \Delta z_i \nonumber
\end{align}

\begin{figure}[h]
  \centering
  \includegraphics[width=0.3\textwidth]{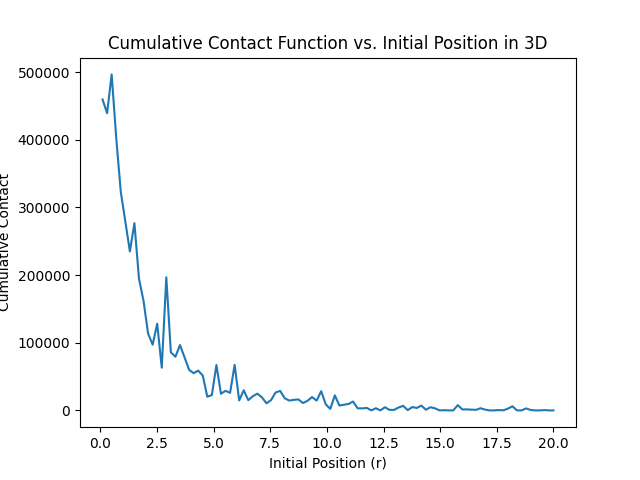}
  \caption{Simulation of cumulative contact counts in three-dimensional random walk model}
  \label{fig:cumurative_03}
\end{figure}

Similarly, the positions $X_2(t)$, $Y_2(t)$, and $Z_2(t)$ of cell 2, initially located at coordinate $r$, were defined as:
\begin{align}
X_2(t) = r + \sum_{i=1}^{t} \Delta x_i \nonumber \\
Y_2(t) = O + \sum_{i=1}^{t} \Delta y_i \nonumber \\
Z_2(t) = O + \sum_{i=1}^{t} \Delta z_i \nonumber
\end{align}
We plotted the contact count $C(r,t)$ with varying initial cell-to-cell distances $r$ (Figure \ref{fig:cumurative_03}). As the initial cell-to-cell distance $r$ increased, the cumulative contact count decreased. The simulation program can be found in Supplement 7.

\begin{figure}[h]
 \centering
 \begin{minipage}[b]{0.46\linewidth}
  \centering
  \includegraphics[width=0.9\textwidth]{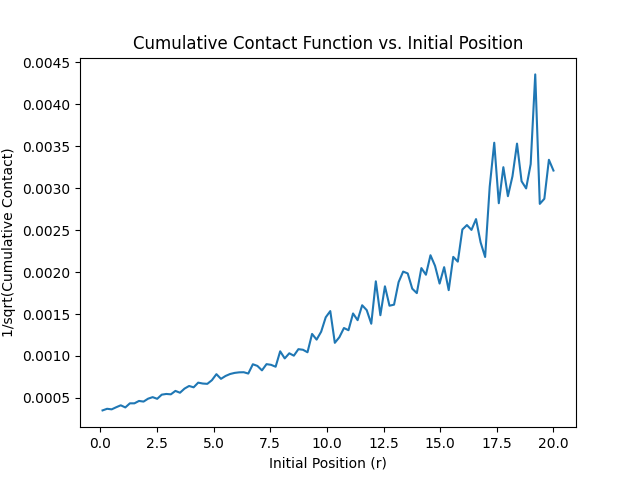}
  \subcaption{The plot of the reciprocal of the square root of the cumulative contact count in a two-dimensional random walk model}
  \label{fig:cumurative_02_inverse}
 \end{minipage}
 \hspace{0.03\linewidth} 
 \begin{minipage}[b]{0.46\linewidth}
  \centering
  \includegraphics[width=0.9\textwidth]{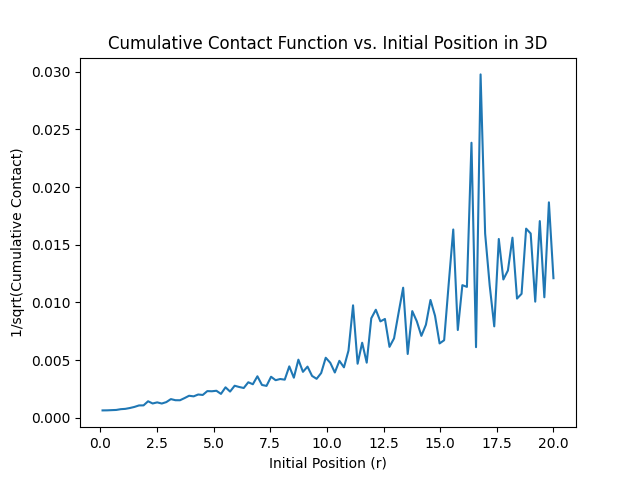}
  \subcaption{The plot of the reciprocal of the square root of the cumulative contact count in a three-dimensional random walk model}
  \label{fig:cumurative_03_inverse}
\end{minipage}
  \caption{The plot of the reciprocal of the square root of the cumulative contact count in a random walk model}
  \label{fig:cumurative_02_03_inverse}
\end{figure}

To investigate the inverse power dependence of distance, the reciprocal of the square root of the cumulative contact function was plotted to assess the presence of a linear relationship. The two-dimensional and three-dimensional simulation programs are provided in Supplement 5 and Supplement 8, respectively. The results of the plots suggested a linear relationship between the initial cell-to-cell distance $r$ and $C(r)$, indicating an inverse square relationship in both two-dimensional and three-dimensional scenarios (see Figure \ref{fig:cumurative_02_03_inverse}).
\[
C(r) \propto \frac{1}{r^2}
\]

%% file: 04_application.tex
Patch images were extracted using whole slide images (WSI) of colorectal cancer. Among images showing the co-presence of cancer and lymphocytes, 286 were randomly selected. Each image then underwent cell detection using AI. Furthermore, we calculated the sum of the composite spatial interaction potential ($\sum \Phi_c$) at every point of a grid-based coordinate system, and investigated the correlation between CLI and $\sum \Phi_c$, hereinafter simply referred to as SIP for simplicity.

\begin{figure}[h]
 \centering
 \begin{minipage}[b]{0.45\linewidth}
  \centering
  \includegraphics[width=0.9\textwidth]{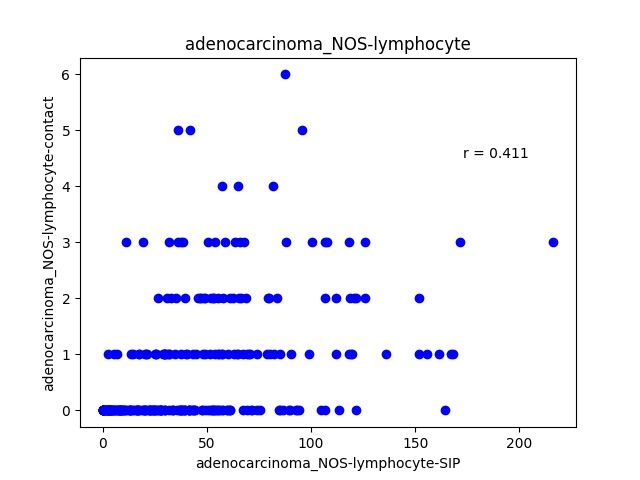}
  \subcaption{Plot of SIP and Contact Numbers Between Cancer Cells and Lymphocytes}
  \label{fig:contact-SIP}
 \end{minipage}
 \hspace{0.03\linewidth} 
 \begin{minipage}[b]{0.45\linewidth}
  \centering
  \includegraphics[width=0.9\textwidth]{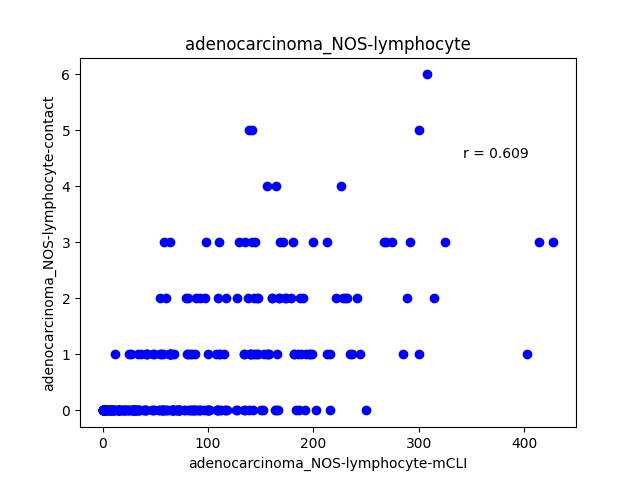}
  \subcaption{Plot of CLI and Contact Numbers Between Cancer Cells and Lymphocytes}
  \label{fig:contact-CLI}
\end{minipage}
  \caption{Relationship Between Contact Numbers, SIP, and CLI in Actual Colorectal Cancer Images}
  \label{fig:contact-SIP-CLI}
\end{figure}

Figure \ref{fig:contact-SIP-CLI} illustrates the relationship between the number of cell contacts, SIP, and CLI in biological images. While a moderate positive correlation between SIP and contact numbers is suggested (Figure \ref{fig:contact-SIP}), a stronger positive correlation was indicated between CLI and contact numbers (Figure \ref{fig:contact-CLI}).

\begin{figure}[h]
 \centering
 \begin{minipage}[b]{0.45\linewidth}
  \centering
  \includegraphics[width=0.9\textwidth]{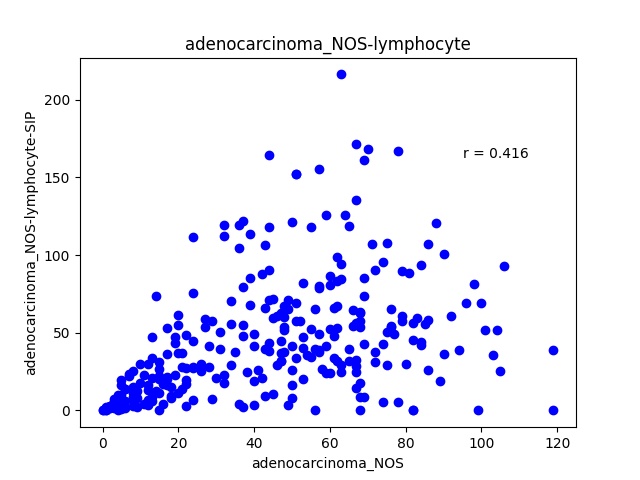}
  \subcaption{Plot of Cancer Cell Count and SIP}
  \label{fig:Ca_no-SIP}
 \end{minipage}
 \hspace{0.03\linewidth} 
 \begin{minipage}[b]{0.45\linewidth}
  \centering
  \includegraphics[width=0.9\textwidth]{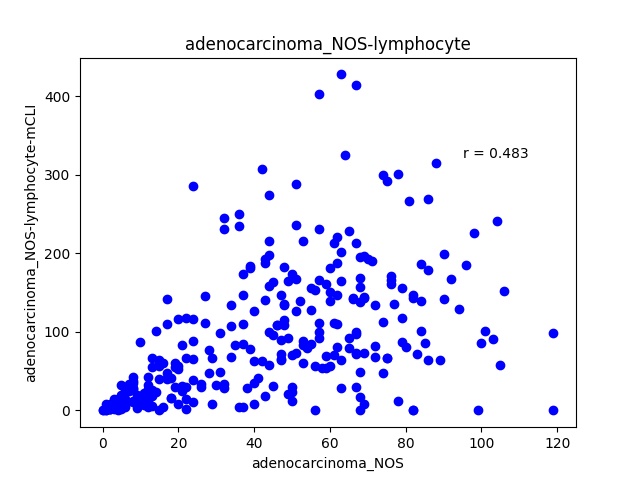}
  \subcaption{Plot of Cancer Cell Count and CLI}
  \label{fig:Ca_no-CLI}
\end{minipage}
  \caption{Relationship Between Cancer Cell Count, SIP, and CLI}
  \label{fig:Ca_no-SIP-CLI}
\end{figure}

\begin{figure}[h]
 \centering
 \begin{minipage}[b]{0.45\linewidth}
  \centering
  \includegraphics[width=0.9\textwidth]{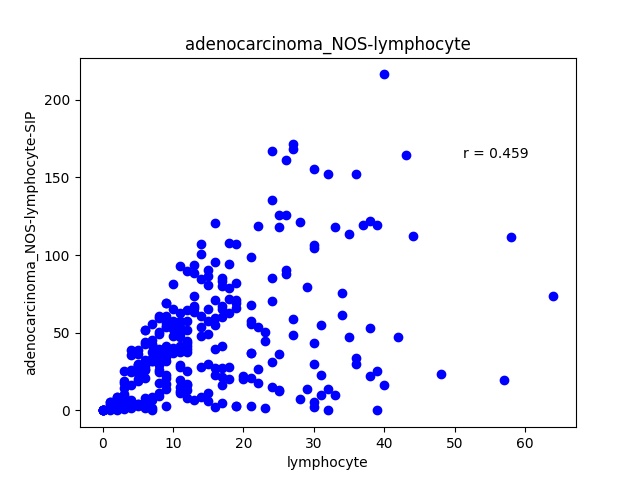}
  \subcaption{Plot of Lymphocyte Count and SIP}
  \label{fig:lym_no-SIP}
 \end{minipage}
 \hspace{0.03\linewidth} 
 \begin{minipage}[b]{0.45\linewidth}
  \centering
  \includegraphics[width=0.9\textwidth]{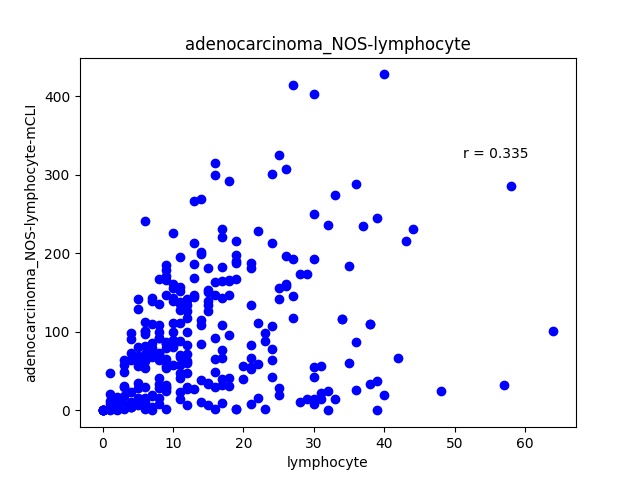}
  \subcaption{Plot of Lymphocyte Count and CLI}
  \label{fig:lym_no-CLI}
\end{minipage}
  \caption{Relationship Between Lymphocyte Count, SIP, and CLI}
  \label{fig:lym_no-SIP-CLI}
\end{figure}

\begin{figure}[h]
  \centering
  \includegraphics[width=0.3\textwidth]{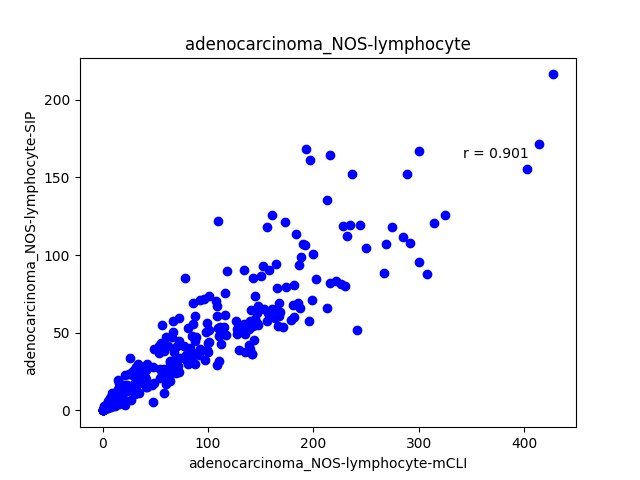}
  \caption{Relationship Between SIP and CLI Between Cancer Cells and Lymphocytes}
  \label{fig:SIP-CLI}
\end{figure}

The relationship between the number of cancer cells and SIP is shown in Figure \ref{fig:Ca_no-SIP}. Areas with a higher number of cancer cells tended to have higher SIP between cancer cells and lymphocytes. The relationship between the number of cancer cells and CLI is presented in Figure \ref{fig:Ca_no-CLI}. Similarly, areas with a higher number of cancer cells showed a tendency for higher CLI between cancer cells and lymphocytes. A similar trend was observed for the relationship between the number of lymphocytes and SIP (Figure \ref{fig:lym_no-SIP}), as well as the relationship between lymphocytes and CLI (Figure \ref{fig:lym_no-CLI}).
Subsequently, when the relationship between SIP and CLI was examined, the two were found to correlate well (Figure \ref{fig:SIP-CLI}).

\begin{figure}[h]
 \centering
 \begin{minipage}[b]{0.46\linewidth}
  \centering
  \includegraphics[width=0.9\linewidth]{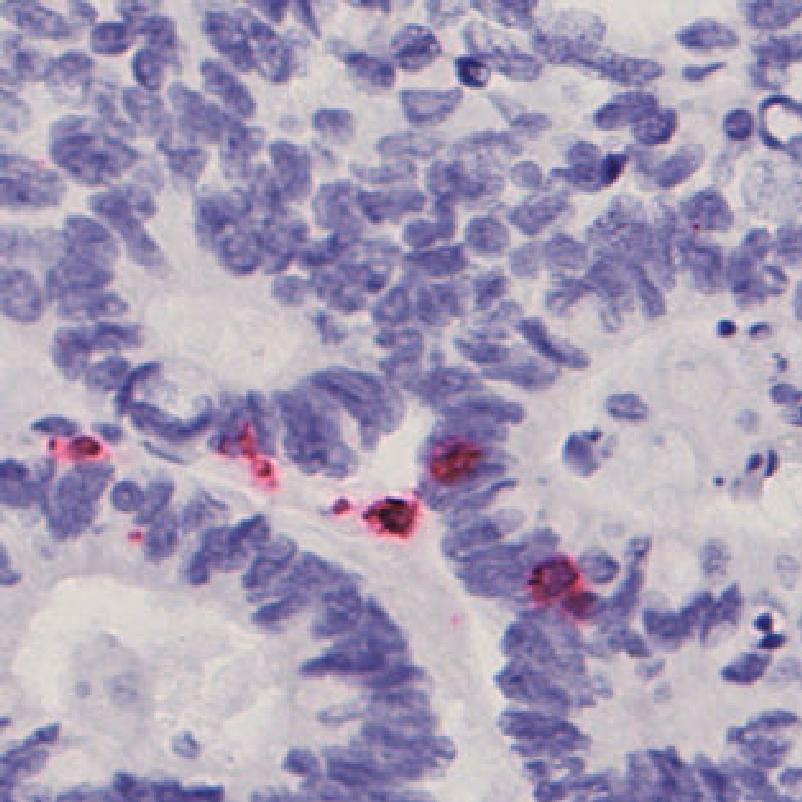}
  \subcaption{Original image}
  \label{fig:orig}
 \end{minipage}
 \hspace{0.03\linewidth} 
 \begin{minipage}[b]{0.46\linewidth}
  \centering
  \includegraphics[width=0.9\linewidth]{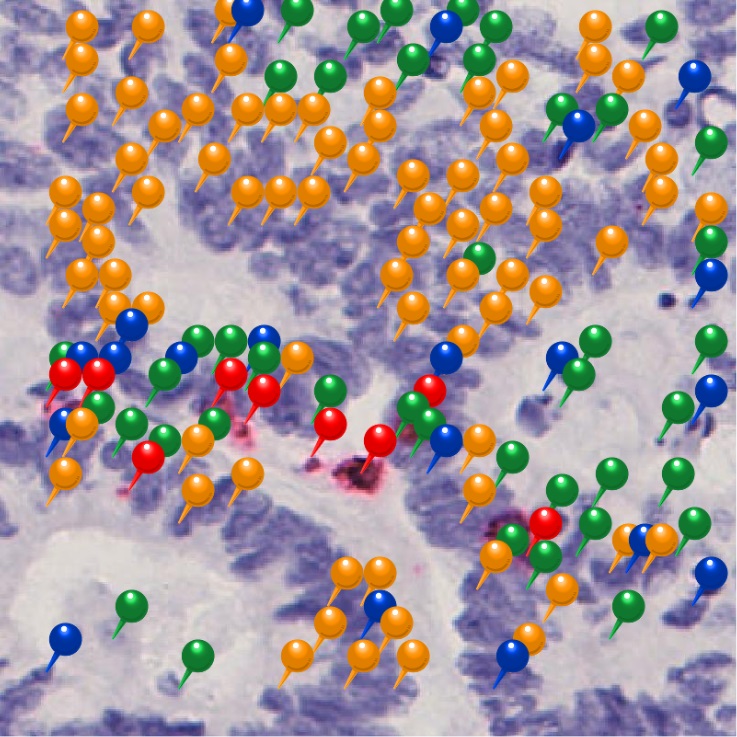}
  \subcaption{Output from AI}
  \label{fig:orig_AIout}
 \end{minipage}\\
 \begin{minipage}[b]{0.46\linewidth}
  \centering
  \includegraphics[width=0.9\linewidth]{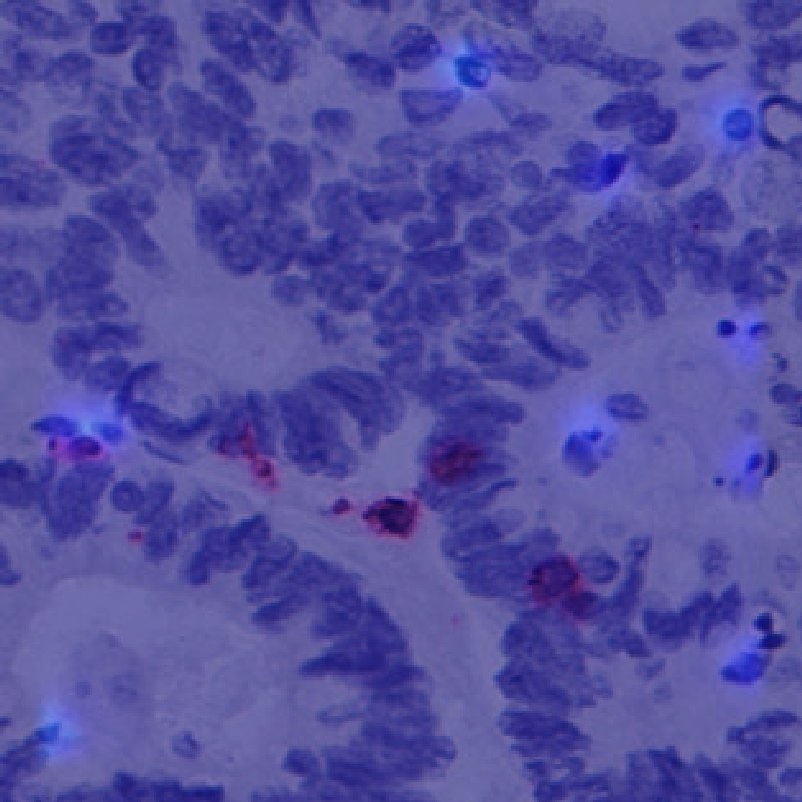}
  \subcaption{SIM for lymphocyte \\3.8 mCLI/mm$^2$}
  \label{fig:CLI1_lymph}
 \end{minipage}
 \hspace{0.03\linewidth} 
 \begin{minipage}[b]{0.46\linewidth}
  \centering
  \includegraphics[width=0.9\linewidth]{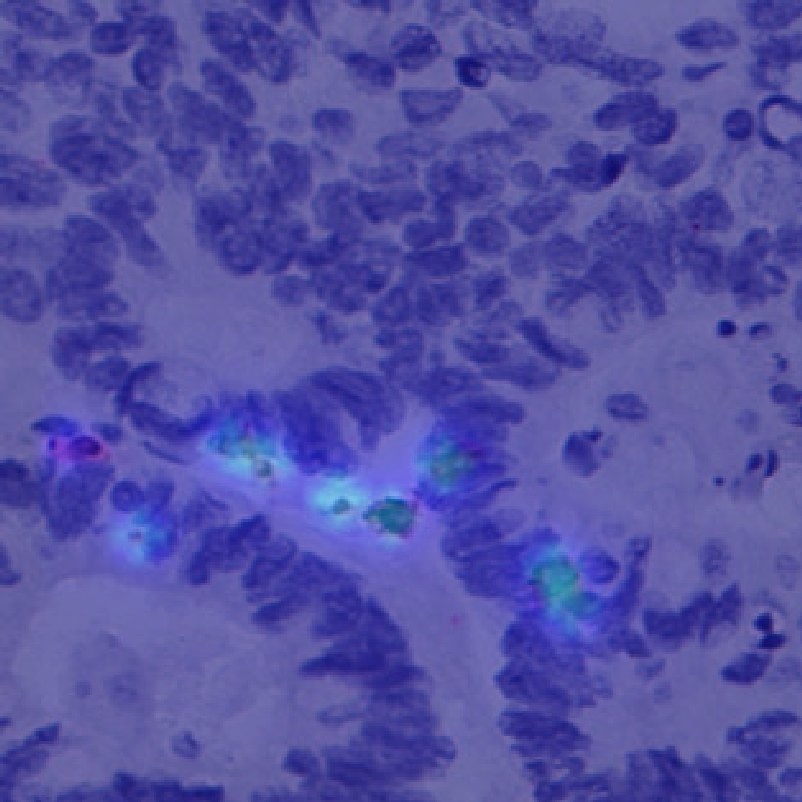}
  \subcaption{SIM for CD8$^+$ lymphocyte \\36.6 mCLI/mm$^2$}
  \label{fig:CLI1_red_m}
 \end{minipage}
 \caption{SIM for single cell type and CLI \\
(a): The original image depicts lymphocyte infiltration surrounding the colorectal cancer. CD8 is labeled in red (Fast Red).
(b): The AI outputs coordinate and labeling information for each cell type. Here, cancer (orange), CD8 (red), other lymphocytes (blue), and other cells such as macrophages and fibroblasts (green) are identified.
(c): Lymphocytes are sparsely distributed, resulting in a low representation in the Spatial Interaction Map (SIM) and a corresponding low CLI value of 3.8.
(d): CD8$^+$ lymphocytes exhibit slight clustering, indicating a relatively higher value in the SIM. The CLI value has also increased to 36.6.}
 \label{fig:CLI1s}
\end{figure}

The spatial interaction potential was visualized using immunohistochemical staining images of colorectal cancer specimens. CD8 was labeled with Fast Red. In the original image (Figure \ref{fig:orig}), cells identified by AI including adenocarcinoma cells (orange), CD8$^+$ lymphocytes (red), other lymphocytes (blue), macrophages, and stromal cells (green) are indicated by pins (Figure \ref{fig:orig_AIout}).
The CLI related to the type of lymphocyte was 3.8 mCLI/mm$^2$, indicating a low value (Figure \ref{fig:CLI1_lymph}).
The CLI of CD8$^+$ lymphocytes in the figure \ref{fig:CLI1_red_m} was 36.6 mCLI/mm$^2$, demonstrating that this method quantitatively detected that these cells were clustered.

\begin{figure}[h]
 \centering
 \begin{minipage}[b]{0.46\linewidth}
  \centering
  \includegraphics[width=0.9\linewidth]{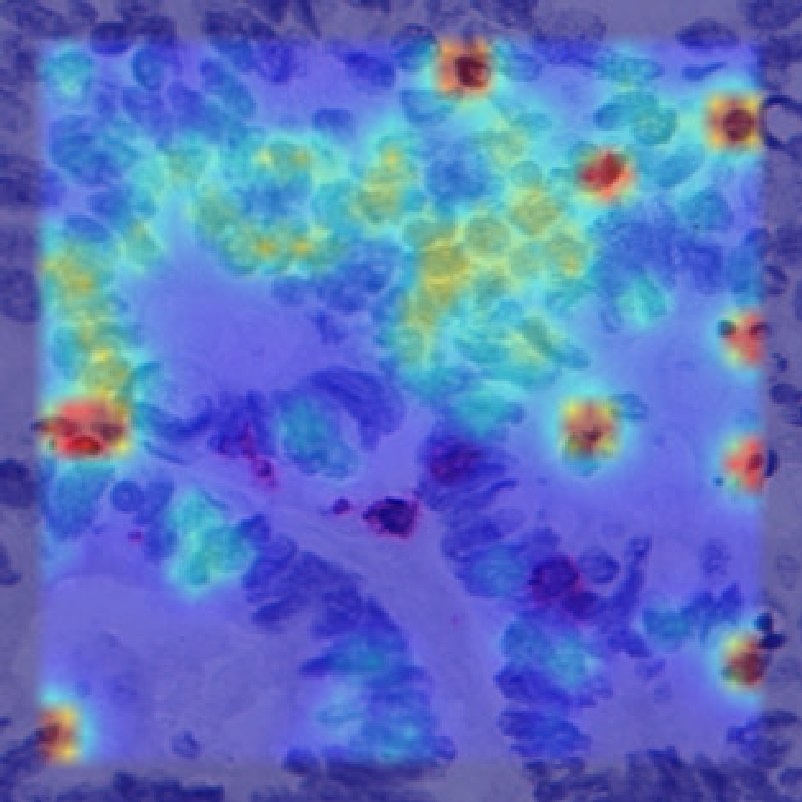}
  \subcaption{SIM for between cancer and lymphocyte \\115.3 mCLI/mm$^2$}
  \label{fig:CLI2_ca_lymph}
 \end{minipage}
 \hspace{0.03\linewidth} 
 \begin{minipage}[b]{0.46\linewidth}
  \centering
  \includegraphics[width=0.9\linewidth]{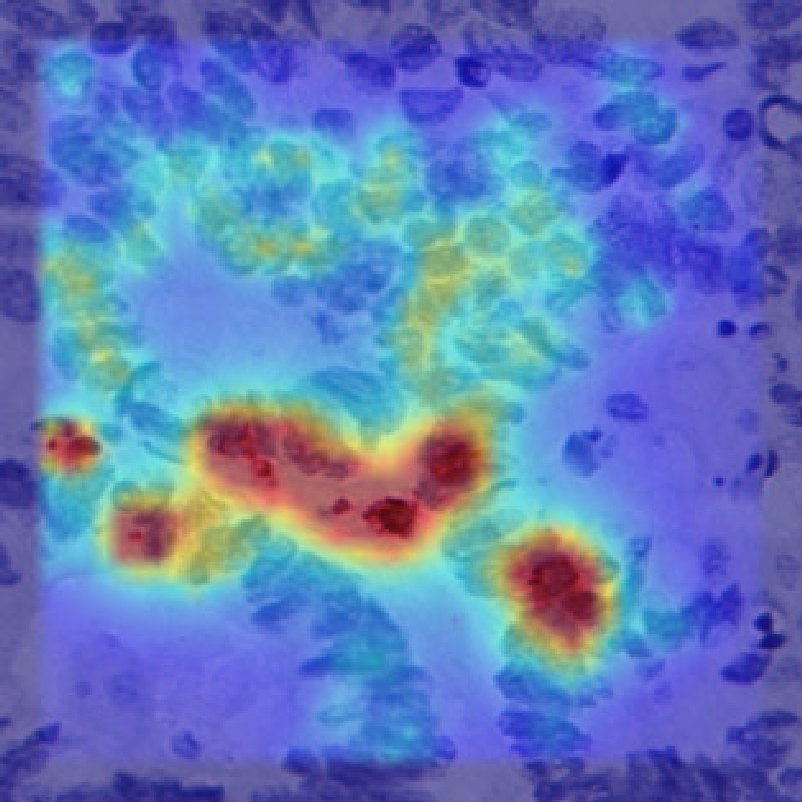}
  \subcaption{SIM for between cancer and CD8$^+$lymphocyte \\250.8 mCLI/mm$^2$}
  \label{fig:CLI2_ca_red_m}
 \end{minipage}\\
 \begin{minipage}[b]{0.46\linewidth}
  \centering
  \includegraphics[width=0.9\linewidth]{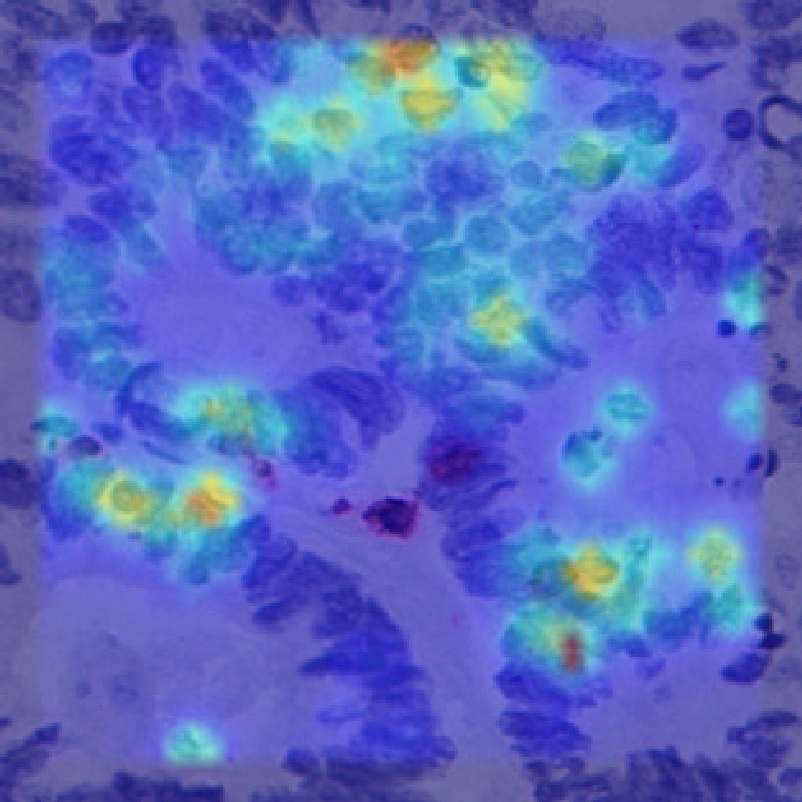}
  \subcaption{SIM between cancer cells, CD8$^+$ lymphocytes, and macrophages: \\32.1 mCLI/mm$^2$}
  \label{fig:CLI3_ca_redm_macro}
 \end{minipage}
 \hspace{0.03\linewidth} 
 \begin{minipage}[b]{0.46\linewidth}
  \centering
  \includegraphics[width=0.9\linewidth]{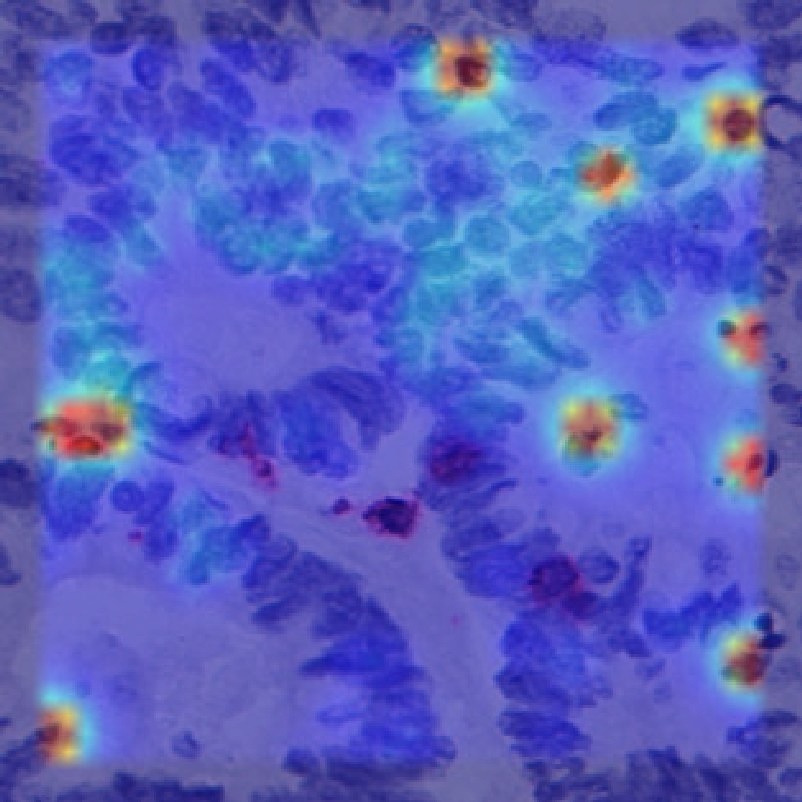}
  \subcaption{SIM between cancer cells, other lymphocytes, and macrophages: \\15.5 mCLI/mm$^2$}
  \label{fig:CLI3_ca_lymph_macro}
 \end{minipage}
  \caption{SIMs between 2 and 3 cells and corresponding CLI\\
  (a): The CLI between cancer cells and lymphocytes remains relatively low due to the scattered distribution of lymphocytes.
(b): The CLI between cancer cells and CD8$^+$ lymphocytes indicates a high value, as CD8$^+$ cells cluster in the vicinity of the cancer.
(c): The Spatial Interaction Map (SIM) between cancer cells and CD8$^+$ lymphocytes, along with macrophages.
(d): The SIM among cancer cells, lymphocytes, and macrophages. The low CLI value is attributed to the scattered distribution of lymphocytes among the three cell types.}
 \label{fig:CLI23s}
\end{figure}

The CLI between two cells and three cells was also investigated. Figure \ref{fig:CLI2_ca_lymph} depicts the spatial interaction map (SIM) between cancer cells and lymphocytes. As the lymphocytes are scattered, the CLI remains relatively moderate at 115.3 mCLI/mm$^2$.
Conversely, Figure \ref{fig:CLI2_ca_red_m} reflects the SIM between cancer cells and CD8+ lymphocytes resulting in a CLI of 250.8 mCLI/mm$^2$, indicating that this method can quantitatively detect dense CD8+ lymphocytes around cancer cells.
Subsequently, the comparison between Figures \ref{fig:CLI3_ca_redm_macro} and \ref{fig:CLI3_ca_lymph_macro} for three-cell interactions was made. While the latter appears to exhibit a higher CLI from the SIM, the former demonstrated a CLI of 32.1 mCLI/mm$^2$, whereas the latter had a value of 15.5 mCLI/mm$^2$. The presence of scattered lymphocytes, forming the combination in the latter case, is thought to be influencing this outcome.

%% file: 05_discussion.tex
In this study, we have developed an innovative and robust approach for analyzing and quantifying cellular interactions on digital pathology images.
By considering the probability of cell presence through the utilization of deep learning and the distances between cells, CLI can assess the accumulation of intercellular interaction based on spatial arrangements. Moreover, the evaluation method for CLI is represented by a simple and intuitively understandable formula. Furthermore, CLI can be generalized to combinations of more than three cell types.

Our method utilizes deep learning to identify cells and determine their probabilities, allowing us to infer the likelihood of a cell's presence in a given location. 
This approach offers a potential solution to restore the quantitativeness of cells that may be overlooked by object-based cell detection methods.
By assuming the probability of a cell's presence, we can probabilistically calculate whether cells in close proximity will come into contact and interact as a result of follwing cell movement.  Using this method, we first simulated the frequency of cell-cell contact within a certain distance over a given period. 
The results showed that, both in two and three dimensions, the cumulative number of contacts is inversely proportional to the square of the initial distance between the cells $r$. This suggests that once the probability of cell presence $p$ and the intercellular distance $r$ are known, it is possible to quantify the contact between cells that leads to CIF $\Psi $. 
This finding aligns with the commonly observed behavior in various physical interactions and diffusion processes where the contact count diminishes proportionally to the square of the distance.
Based on the simulation results, we defined the CIF $ \Psi$ as follows:
\[
\Psi = K_c \frac{p_1 p_2}{r^2}
\]
In dense cellular environments, it is expected that cells within the same locality will engage in complex interactions. Therefore, we further summed the CIF of individual pairs and defined the CLI as follows:
\[
\Omega = K_c \sum_{i\in A} \sum_{j\in B} \frac{{p}_i {p}_j}{r^2}
\]
CLI $\Omega$ is a quantitative method for analyzing cell interactions that is derived by probabilistically considering spaciotemporal cell-cell contacts. The approach of deriving the probability of cell presence using deep neural networks offers significant benefits. By assigning the evaluation of cell presence probability to deep learning-based cell detection before calculating CLI, management becomes simpler. Ensuring that the learning dataset is sufficiently large and accurately reflects the population guarantees the precision of cell detection. In essence, quantifying cellular co-localization largely depends on the accuracy of deep learning, making the formula for calculating CLI exceedingly simple, clear, and robust. This allows developers to focus on creating algorithms for accurate cell detection and preparing datasets.
Another advantage is its immediate applicability to commonly used specimens, such as standard HE staining and immunostaining. This is particularly critical in an era where expensive analytical methods, such as multiplex analysis and spatial omics analysis, are becoming more common. This approach is promising, given the technological advancements that enable cell detection in unstained \citep{rivenson2019virtual} or faded specimens \citep{sun2023color}.

We explain the reason for using CLI instead of directly utilizing the spatial interaction potential (SIP) derived for CLI. CLI reflects the distance factor more strongly than the sum of SIP. Furthermore, CLI requires simpler calculations and has a lower computational load.
We demonstrated that a correlation exists between the sum of the composite SIP ($\Phi_c$) and CLI.
The composite SIP $\Phi_c$ and CLI are dependent on common parameters such as the existence probability and the interaction constant $K_c$. Thus, a correlation between $\Phi_c$ and CLI can be anticipated, although they are not identical. The CLI is based on the probabilities of the presence of Cell 1 and Cell 2 and their distance, which directly influences the correlation with the cellular distribution. For example, in areas where Cells 1 and 2 are densely present, the CLI is higher. On the contrary, in areas where Cells 1 and 2 are located further apart, the CLI is lower. The impact of cellular distribution on the correlation between the composite potential $\Phi_c$ and CLI varies with the probabilities of Cell 1 and Cell 2's presence, the interaction constant $K_c$, and the characteristics of the cellular distribution. Generally, in areas with a denser cellular distribution, the values in CLI, compared to $\Phi_c$, are higher. However, when the cellular distribution is uniform, the values of CLI, in comparison to $\Phi_c$, do not drastically increase.
Therefore, composite SIP and CLI differ slightly in their biological implications. In SIP, the potential for each cell is calculated without the distances between cells affecting the final sum, meaning SIP is an index that simply scales with the number of cells. In contrast, CLI is inversely proportional to the square of the distance $r$ between cells, which means that for the same number of cells, the value increases when cells are closer together and clustered. Therefore, CLI not only reflects the number of cells but also their clustering. 
The reason why CLI showed a stronger correlation with contact numbers compared to SIP might be attributed to these factors. 
The computation of SIP requires calculating $\Phi_c$ for each combination of cells at every point in a grid-based coordinate system, which involves a significant amount of computation. On the other hand, CLI only requires calculating cell interactions based on the number of detected cells, thus offering a considerable computational advantage.

However, composite SIP is valuable for visualization. We refer to it as the spatial interaction map (SIM). 
SIM is a tool that allows for the visualization of the composite SIP $\Phi_c$ for specific cell combinations. In Section 4, the application of SIM on actual biological images demonstrated its effectiveness. The interactions between cells were clearly depicted as a colormap, confirming its usefulness for visual understanding.
Although the impressions received from the visualized SIP do not fully correlate with CLI, it is considered to have a certain level of usefulness for the image-based evaluation of CLI.

Next, we will discuss the limitations in sequence.
Regarding the definition of CLI, there indeed are concerns about treating the classification probability of cells as their existence probability. As mentioned in Section 2, to appropriately handle such probabilities as existence probabilities, many factors dependent on the probability $q$ need to be considered.
Specifically, the following steps are necessary: first, an evaluation at the patch level to assess the accuracy of the assigned probabilities, second, an assessment at the count level to ensure the proper creation of pin positions and counts, and third, an evaluation of the annotation to ensure that the classification reflects the cell population of the parent population appropriately.
Through these steps, the evaluation determines whether the probabilities in the "classification space" used by a specific AI can be regarded as the "presence probability" of cells on actual pathology slides. To address concerns related to these probabilities $q$, an appropriate approach involves combining data collection and sampling methodologies, statistical techniques, annotation plans, machine learning, and domain expertise to create a cell dataset that appropriately reflects the parent population. However, as these approaches extend beyond the scope of this paper, they are deferred to another opportunity.

There are concerns regarding the validity of assuming that cell interactions are inversely proportional to the square of the distance. In general, it is common for interactions between cells to be influenced by the square of the distance. This is attributed to the observations of effects proportional to the square of the distance in many physical phenomena such as Coulomb forces, gravity, and diffusion. Such distance-squared-influenced interactions are applicable to numerous physical laws and phenomena, and they are believed to be applicable in modeling cell-cell interactions.

For the current study, the use of the uniform distribution in the random walk model for cell movement raises valid concerns about whether it provides a sufficiently accurate simulation. The persistent random walk model assumes that particles exhibit random movement while maintaining a persistent trend in a specific direction and has been widely utilized in the analysis of cell movement patterns \citep{othmer1988models, dickinson1993optimal}. Subsequent research has further modified and extended these models. For instance, \cite{maree2006polarization} constructed a model considering diffusion of specific substances in the cytoplasm and membrane to account for nonlinear relationships in the movement of epidermal cells. Additionally, \cite{schluter2012computational} modeled the influence of the extracellular matrix on cell movement. Various other models of cell interactions have also been developed \citep{danuser2013mathematical}. Likely, the extent to which cell-cell contact frequencies and interactions are influenced by distance depends on the specific configuration of the model. One of the limitations of simulation studies is their dependence on the specific setup of the model. It would require further research, including prognostic analyses, to determine the biological validity of the CLI developed in this study.

For the practical implementation of CLI, it is important to determine the constant $K_c$. We considered an ideal scenario where cells with a probability of 1 are densely packed in order to determine the constants. There could be other variations to consider regarding the size and arrangement of cells, and the choice of 100 cells is somewhat arbitrary. However, this choice was made considering situations where relatively small cells, such as lymphocytes, tend to cluster, and it is believed to have some biological plausibility. Moreover, it is considered that this metric could be standardized to some extent for not only a single cell type but also among multiple cell types.

\newpage

In this study, we have proposed the CLI as a novel quantitative method for intercellular interactions. The spatial interactions utilized in this work are purely geometric, and further investigation is required to ascertain the biological utility of the CLI as presented in this study.

The following are descriptions of the advantages of the proposed CLI:
\begin{itemize}
\item A novel approach to handling cell presence probabilities: Under the constraints of fixed pathology specimens, CLI offers a new method that considers the probability of cell presence through the utilization of deep learning.
\item Consideration of spatial arrangement: By taking into account the distances between cells, CLI can assess the accumulation of intercellular interaction potentials based on spatial arrangements.
\item Intuitive evaluation method: The evaluation method of CLI is represented by a simple and intuitively understandable formula.
\item Generalization to the evaluation of co-localization among multiple cells: CLI can be applied to the evaluation of co-localization among combinations of more than three cell types, thus possessing features not present in conventional methods for analyzing intercellular co-localization. It enables the comparison of cell interactions by evaluating different combinations of cells.
\end{itemize}

The proposed CLI is an objective and reproducible new quantitative method for intercellular interactions. In actual pathological tissue images, CLI demonstrates high performance. Further investigation is expected to elucidate the biological significance of CLI.

%% file: 06_supplement.tex
\section*{Supplement 1}\label{code_1d_final_position}

{\scriptsize
\begin{verbatim}
import numpy as np
import matplotlib.pyplot as plt

T = 1000  
delta_x_max = 1.0  
num_simulations = 1000  

x = np.zeros((num_simulations, T+1))

for i in range(num_simulations):
    for t in range(T):
        delta_x = np.random.uniform(-delta_x_max, delta_x_max)
        x[i, t+1] = x[i, t] + delta_x

final_positions = x[:, -1]

plt.hist(final_positions, bins=30, density=True, alpha=0.6, color='b',
           label='Position at t=T')
plt.xlabel('Position (x)')
plt.ylabel('Probability Density')
plt.title('Distribution of Final Positions')
plt.legend()
plt.show()
\end{verbatim}
}

\section*{Supplement 2}\label{code_1d_cummrative_contact}

{\scriptsize
\begin{verbatim}
T = 1000 
d0 = 1.0  
num_simulations = 100  
num_positions = 100  

cumulative_contacts = np.zeros(num_positions)

r_values = np.linspace(0.1, 20.0, num_positions)

for idx, r in enumerate(r_values):
    cumulative_contact = np.zeros(T)
    
    for _ in range(num_simulations):
        
        X1 = 0.0
        X2 = r
        
        contact_count = 0  
        
        for t in range(T):
            delta_x1 = np.random.uniform(-0.5, 0.5)
            X1 += delta_x1
            
            delta_x2 = np.random.uniform(-0.5, 0.5)
            X2 += delta_x2
            
            distance = np.abs(X2 - X1)
            
            if distance < d0:
                contact_count += 1
            
            cumulative_contact[t] += contact_count
    
    cumulative_contacts[idx] = np.mean(cumulative_contact) / num_simulations

plt.plot(r_values, cumulative_contacts)
plt.xlabel('Initial Position (r)')
plt.ylabel('Cumulative Contact')
plt.title('Cumulative Contact Function vs. Initial Position')
plt.show()
\end{verbatim}
}

\section*{Supplement 3}\label{code_2d_final_position}

{\scriptsize
\begin{verbatim}
from mpl_toolkits.mplot3d import Axes3D

T = 1000  
delta_x_max = 1.0 
delta_y_max = 1.0 
num_simulations = 1000  

x = np.zeros((num_simulations, T + 1))
y = np.zeros((num_simulations, T + 1))

for i in range(num_simulations):
    for t in range(T):
        delta_x = np.random.uniform(-delta_x_max, delta_x_max)
        delta_y = np.random.uniform(-delta_y_max, delta_y_max)
        x[i, t + 1] = x[i, t] + delta_x
        y[i, t + 1] = y[i, t] + delta_y

final_positions_x = x[:, -1]
final_positions_y = y[:, -1]

fig = plt.figure()
ax = fig.add_subplot(111, projection='3d')
hist, xedges, yedges = np.histogram2d(final_positions_x, final_positions_y, bins=30)
xpos, ypos = np.meshgrid(xedges[:-1] + xedges[1:], yedges[:-1] + yedges[1:])
xpos = xpos.flatten() / 2.
ypos = ypos.flatten() / 2.
zpos = np.zeros_like(xpos)
dx = dy = 0.5 * (xedges[1] - xedges[0])
dz = hist.flatten()
ax.bar3d(xpos, ypos, zpos, dx, dy, dz, color='b', zsort='average')
ax.set_xlabel('Position (x)')
ax.set_ylabel('Position (y)')
ax.set_zlabel('Frequency')
ax.set_title('3D Histogram of Final Positions in 2D Random Walk')
plt.show()
\end{verbatim}
}

\section*{Supplement 4}\label{code_2d_cummrative_contact}

{\scriptsize
\begin{verbatim}
T = 1000  
d0 = 1.0  
num_simulations = 100  
num_positions = 100

cumulative_contacts = np.zeros(num_positions)

r_values = np.linspace(0.1, 20.0, num_positions)

for idx, r in enumerate(r_values):
    for _ in range(num_simulations):
       
        X1 = np.array([0.0, 0.0])
        X2 = np.array([r, 0.0])
        
        contact_count = 0  
         
        for t in range(T):
            delta_x1, delta_y1 = np.random.uniform(-0.5, 0.5, size=2)
            delta_x2, delta_y2 = np.random.uniform(-0.5, 0.5, size=2)
            X1 += np.array([delta_x1, delta_y1])
            X2 += np.array([delta_x2, delta_y2])
            
            distance = np.linalg.norm(X2 - X1)
            
            if distance < d0:
                contact_count += 1
            
            cumulative_contacts[idx] += contact_count

plt.plot(r_values, cumulative_contacts)
plt.xlabel('Initial Position (r)')
plt.ylabel('Cumulative Contact')
plt.title('Cumulative Contact Function vs. Initial Position')
plt.show()
\end{verbatim}
}

\section*{Supplement 5}\label{code_2d_cummrative_contact_inverse_square}

{\scriptsize
\begin{verbatim}
T = 1000  
d0 = 1.0  
num_simulations = 500  
num_positions = 100  

cumulative_contacts = np.zeros(num_positions)

r_values = np.linspace(0.1, 20.0, num_positions)

for idx, r in enumerate(r_values):
    for _ in range(num_simulations):
        
        X1 = np.array([0.0, 0.0])
        X2 = np.array([r, 0.0])
        
        contact_count = 0 
        
        for t in range(T):
            
            delta_x1, delta_y1 = np.random.uniform(-0.5, 0.5, size=2)
            delta_x2, delta_y2 = np.random.uniform(-0.5, 0.5, size=2)
            X1 += np.array([delta_x1, delta_y1])
            X2 += np.array([delta_x2, delta_y2])
            
            distance = np.linalg.norm(X2 - X1)
            
            if distance < d0:
                contact_count += 1
            
            cumulative_contacts[idx] += contact_count

results = np.where(cumulative_contacts != 0, 1/np.sqrt(cumulative_contacts), 0)

plt.plot(r_values, results)
plt.xlabel('Initial Position (r)')
plt.ylabel('1/sqrt(Cumulative Contact)')
plt.title('Cumulative Contact Function vs. Initial Position')
plt.show()
\end{verbatim}
}

\section*{Supplement 6}\label{code_3d_final_position}

{\scriptsize
\begin{verbatim}
from mpl_toolkits.mplot3d import Axes3D

T = 1000  
delta_x_max = 1.0  
delta_y_max = 1.0  
delta_z_max = 1.0  
num_simulations = 1000  

x = np.zeros((num_simulations, T + 1))
y = np.zeros((num_simulations, T + 1))
z = np.zeros((num_simulations, T + 1))

for i in range(num_simulations):
    for t in range(T):
        delta_x = np.random.uniform(-delta_x_max, delta_x_max)
        delta_y = np.random.uniform(-delta_y_max, delta_y_max)
        delta_z = np.random.uniform(-delta_z_max, delta_z_max)
        x[i, t + 1] = x[i, t] + delta_x
        y[i, t + 1] = y[i, t] + delta_y
        z[i, t + 1] = z[i, t] + delta_z

final_positions_x = x[:, -1]
final_positions_y = y[:, -1]
final_positions_z = z[:, -1]

fig = plt.figure()
ax = fig.add_subplot(111, projection='3d')
ax.scatter(final_positions_x, final_positions_y, final_positions_z,
                c='b', marker='o')
ax.set_xlabel('Position (x)')
ax.set_ylabel('Position (y)')
ax.set_zlabel('Position (z)')
ax.set_title('Final Positions in 3D Random Walk')
plt.show()
\end{verbatim}
}

\section*{Supplement 7}\label{code_3d_cummrative_contact}

{\scriptsize
\begin{verbatim}
T = 1000  
d0 = 1.0  
num_simulations = 100  
num_positions = 100 

cumulative_contacts = np.zeros(num_positions)

r_values = np.linspace(0.1, 20.0, num_positions)

for idx, r in enumerate(r_values):
    for _ in range(num_simulations):
        
        X1 = np.array([0.0, 0.0, 0.0])  
        X2 = np.array([r, 0.0, 0.0])  
        
        contact_count = 0 
         
        for t in range(T):
            
            delta_x1, delta_y1, delta_z1 = np.random.uniform(-0.5, 0.5, size=3)  
            delta_x2, delta_y2, delta_z2 = np.random.uniform(-0.5, 0.5, size=3)  
            X1 += np.array([delta_x1, delta_y1, delta_z1])
            X2 += np.array([delta_x2, delta_y2, delta_z2])
            
            distance = np.linalg.norm(X2 - X1)
            
            if distance < d0:
                contact_count += 1
            
            cumulative_contacts[idx] += contact_count

plt.plot(r_values, cumulative_contacts)
plt.xlabel('Initial Position (r)')
plt.ylabel('Cumulative Contact')
plt.title('Cumulative Contact Function vs. Initial Position in 3D')
plt.show()
\end{verbatim}
}

\section*{Supplement 8}\label{code_3d_cummrative_contact_inverse_square}

{\scriptsize
\begin{verbatim}
T = 1000  
d0 = 1.0  
num_simulations = 500  
num_positions = 100  

cumulative_contacts = np.zeros(num_positions)

r_values = np.linspace(0.1, 20.0, num_positions)

for idx, r in enumerate(r_values):
    for _ in range(num_simulations):
        
        X1 = np.array([0.0, 0.0, 0.0])  
        X2 = np.array([r, 0.0, 0.0])  
        
        contact_count = 0  
        
        for t in range(T):
            
            delta_x1, delta_y1, delta_z1 = np.random.uniform(-0.5, 0.5, size=3)  
            delta_x2, delta_y2, delta_z2 = np.random.uniform(-0.5, 0.5, size=3)  
            X1 += np.array([delta_x1, delta_y1, delta_z1])
            X2 += np.array([delta_x2, delta_y2, delta_z2])
            
            distance = np.linalg.norm(X2 - X1)
            
            if distance < d0:
                contact_count += 1
            
            cumulative_contacts[idx] += contact_count

results = np.where(cumulative_contacts != 0, 1/np.sqrt(cumulative_contacts), 0)

plt.plot(r_values, results)
plt.xlabel('Initial Position (r)')
plt.ylabel('1/sqrt(Cumulative Contact)')
plt.title('Cumulative Contact Function vs. Initial Position in 3D')
plt.show()
\end{verbatim}
}